\documentclass{vgtc}                          % final (conference style)
\graphicspath{{figures/}{pictures/}{images/}{./}} % where to search for the images

\usepackage{times}                     % we use Times as the main font
\usepackage{float}

\usepackage{tabu}                      % only used for the table example
\usepackage{booktabs}                  % only used for the table example
\usepackage{lipsum}                    % used to generate placeholder text
\usepackage{mwe}                       % used to generate placeholder figures

\usepackage{mathptmx}                  % use matching math font

\onlineid{0}

\vgtccategory{Research}

\vgtcinsertpkg

\title{ReaWristic: Remote Touch Sensation to Fingers
from a Wristband via Visually Augmented Electro-Tactile Feedback}

\author{Yudai Tanaka\thanks{e-mail: yudaitanaka@uchicago.edu} %
\and Neil Weiss%
\and Robert Cole Bolger-Cruz
\and Jess Hartcher-O'Brien
\and Brendan Flynn
\and Roger Boldu
\and Nicholas Colonnese
}
\affiliation{\scriptsize Reality Labs Research, Meta}

\teaser{
  \centering
  \includegraphics[width=\linewidth]{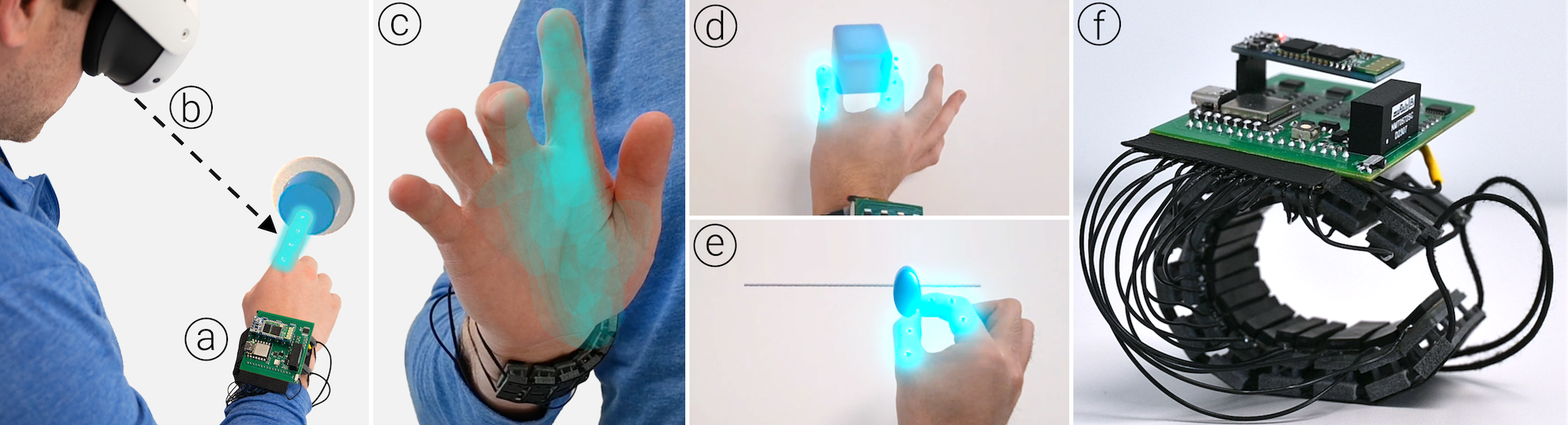}
  \vspace{-20 pt}
  \caption{We present a technique to provide tactile feedback on the fingers from a wristband wearable. We achieve this through (a) electrically stimulating the wrist, and (b) rendering a visual effect in mixed reality (MR) to enhance its perceived localization. (c) With our approach, the user perceives about 40$\sim$50\% of the sensations in a target region (the overlayed blue region represents the area of perceived sensation from Study 2). This enables (d) tactile feedback in grabbing virtual objects or (e) tactile detents in a virtual slider. (f) Unlike most electro-tactile approaches, our device dispenses with gel electrodes, and instead, employs high-biocompatibility elastomer electrodes, making it easy to put on \& take off.}
  \label{fig:teaser}
}

\abstract{
    We present a technique for providing remote tactile feedback to the thumb and index finger via a wristband device. This enables haptics for touch and pinch interactions in mixed reality (MR) while keeping the user’s hand entirely free. We achieve this through a novel cross-modal stimulation, which we term \textit{visually augmented electro-tactile feedback}. This consists of (1) electrically stimulating the nerves that innervate the targeted fingers using our wristband device; and (2) concurrently, visually augmenting the targeted finger in MR to steer the perceived sensation to the desired location. In our psychophysics study, we found that our approach provides tactile perception akin to tapping and, even from the wrist, it is capable of delivering the sensation to the targeted fingers with $\sim$50\% of sensation occurring in the thumb and $\sim$40\% of sensation occurring in the index finger. These results on localizability are unprecedented compared to electro-tactile feedback alone or any prior work for creating sensations in the hand with devices worn on the wrist/arm. Moreover, unlike conventional electro-tactile techniques, our wristband dispenses with gel electrodes. Instead, it incorporates custom-made elastomer-based dry electrodes and a stimulation waveform designed for the electrodes, ensuring the practicality of the device beyond laboratory settings. Lastly, we evaluated the haptic realism of our approach in mixed reality and elicited qualitative feedback from users. Participants preferred our approach to a baseline vibrotactile wrist-worn device.
} % end of abstract


\begin{document}


\firstsection{Introduction}

\maketitle

%% \section{Introduction} %for journal use above \firstsection{..} instead
Today’s consumer headsets (e.g., Meta Quest) have shown that mixed reality (MR) technology has reached some maturity in audiovisual \& tracking aspects. On the contrary, haptic feedback in MR is lagged behind, i.e., users typically do not feel what’s expected in their fingers upon manipulating UIs in MR. A bottleneck for MR haptics is that, while audiovisual \& tracking can be built into the headset itself, to feel feedback in their hand, users have to wear an additional device on their hand. For instance, many haptic devices take the form of a glove. These are undesirable for MR as they constantly cover up the user’s hand, impeding their daily activities, and even MR experience itself, when interactions include manipulating real-world objects.

As such, many posit that wrist-worn haptic devices form a reasonable solution to MR haptics \cite{carcedo2016hapticolor, gupta2017hapticclench, pezent2019tasbi}: they keep the hand entirely unencumbered, and wristbands are the de facto user preferrable form factor for wearables (e.g., smartwatches). However, moving the device to the wrist means that they can only provide feedback to the wrist, not the hand/fingers. Naturally, most wrist-worn haptic devices are limited to vibrotactile \cite{carcedo2016hapticolor} or squeezing \cite{pezent2019tasbi} feedback at the wrist. This unfortunately makes haptics from wristband devices unintuitive and limits their realism, i.e., users only feel sensations at the wrist even though most haptic interactions involve touching or grabbing UIs with their fingers.

An emerging approach is to apply electrical pulses to the nerves at the wrist, evoking tactile sensations in the regions of the hand they supply \cite{ogihara2023multi, pena2021channel}. Unfortunately, while the electro-tactile feedback can provide sensations across the fingers and palm, it falls short in creating adequately localized touch perception for targeting specific fingers, such as the index. Furthermore, the practicality of these devices is limited; they rely on gelled electrodes similar to ones used in neuroscience, which are challenging to attach and detach due to the adhesiveness and must be replaced frequently.

We bridge the gap between wrist-worn devices and electro-tactile feedback, converging them into a practical solution for tactile feedback in MR. Our approach creates sensations in the thumb \& index finger via electrical stimulation at the wrist. This supports prevalent MR interactions such as button-pressing and pinch-grabbing. A core innovation is our cross-modal illusion that enhances the perceived performance of electro-tactile feedback in generating sensations within the targeted fingers---\textit{we term this visually augmented electro-tactile feedback} (\cref{fig:teaser}). In our illusion, an MR application applies a visual effect that highlights the targeted finger (e.g., the index), while a wrist-worn device delivers electro-tactile feedback to the hand, targeting the same finger. As a result, our system provides tactile feedback similar to “tapping”, with an unprecedented rate of it perceived within the targeted region ($\sim$40\% for the index; $\sim$50\% for the thumb), which we found in our Study 1\&2. Notably, the device that applies our electro-tactile feedback is entirely different from most stimulation devices that use sticky gelled electrodes. Instead, our device features custom-made dry electrodes made of conductive elastomer, ensuring the wearability and durability. This frees our device from the adhesive electrodes that require frequent replacement. Users can simply put on and take off our device as they would any other wristband. Finally, in our Study 3, we elicited qualitative feedback from users on the haptic realism, where they preferred our device to a baseline vibrotactile wristband.

\textbf{Contribution and Benefits.} Our key contribution is a novel cross-modal illusion that visually augments electro-tactile feedback, enabling a wristband device to provide tactile perception to the fingers. The benefits of our approach include: (1) achieving state-of-the-art localization of tactile feedback in the fingers, surpassing any existing wrist-worn devices, including those using electro-tactile feedback alone; (2) supporting prevalent haptic interactions in MR, such as button-pressing and pinch-grabbing using the thumb and index finger; and (3) drastically improving the wearability \& durability of electro-tactile devices by employing dry elastomer electrodes to dispense standard adhesive gel electrodes.

\section{Backgound and Related Work}
Our work is built on: (1) haptic devices designed for mixed reality; (2) wrist-worn haptic devices; and (3) referred sensations---haptic sensations perceived in a location different from the stimulus location, specifically those achieved by electro-tactile feedback.
\vspace{-5 pt}
\subsection{Mixed Reality Haptics Needs to Keep Hands Free}
In Mixed Reality (MR), users interact with both virtual \& physical objects. Thus, wearable haptics for MR should keep their hands free, preserving the ability to feel \& manipulate physical objects while adding haptics to virtual UIs; traditional haptic devices (e.g., haptic gloves \cite{noauthor_cybergrasp_nodate, varga_haptx_nodate}) cover the hand and are not suitable for MR.

\textbf{Thin actuators.} One approach is to engineer very thin actuators through which users can feel physical materials, e.g., a thin film of electrodes that stimulates the fingerpad \cite{withana2018tacttoo}, or a PVC tube wrapped around the fingerpad that flows liquids for feedback \cite{han2018hydroring}. 

\textbf{Foldable actuators.} Another approach is to design foldable mechanical structures that cover the fingerpad/palm only during haptic feedback but otherwise sit on a different location to keep the hands/fingers free, e.g., fingerpad-to-nail \cite{teng2021touch} or palm-to-arm \cite{kovacs2020haptic}.

\textbf{Relocated Actuators.} Alternatively, one can simply present haptics to another body part while representing the sensation for the fingerpad or the hand, e.g., nail-mounted actuators for the fingerpad sensations \cite{ando2002smartfinger, preechayasomboon2021haplets} or mechanical pressure to the forearm \cite{moriyama2022wearable}.

While these strategies enhance the user’s ability to feel and manipulate physical objects in MR, each has its own limitation: thin actuators still degrade the user’s tactile acuity compared to bare skin \cite{nittala2019like}; foldable actuators involve latency due to mechanical actuation (340ms for \cite{kovacs2020haptic} and 92ms for \cite{teng2021touch}); and relocated actuators compromise the haptic realism of virtual UIs by presenting feedback to parts other than the fingerpad or hand. Moreover, they require hardware around the finger or forearm. Consequently, from the user’s perspective, it feels as if they must wear these highly specialized devices solely for haptic feedback, which diminishes their social acceptability. As such, many have turned to haptic devices worn on the wrist---a primary location for wearables (e.g., smartwatches) with social acceptability.
\vspace{-5 pt}
\subsection{Wrist-Worn Haptic Devices}
A critical advantage of wrist-worn devices is their potential to merge into standard wearables already on users’ wrists, e.g., smartwatches. Thus, many have explored vibrotactile feedback at the back of the watch, similar to mainstream smartwatches (e.g., Apple Watch), from a single vibrator for notifications \cite{pasquero2011haptic} to multiple vibrators for spatial cues \cite{chen2008tactor, liao2016edgevib}. Also, envisioning the watch strap form factor, many arrange actuators around the wrist, e.g., vibration actuators around the wrist for eyes-free interactions or pathfinding \cite{carcedo2016hapticolor, salazar2018path}. To expand the range of sensations, researchers also turned to mechanical pressures around the wrist (i.e., squeezing), using inflatable straps \cite{pohl2017squeezeback} or even shape memory alloys \cite{gupta2017hapticclench}. Recently, Pezent et al. developed Tasbi, a wristband device for both vibrotactile \& squeezing feedback using motorized strings connected to vibration tactors \cite{pezent2019tasbi}. Notably, Tasbi invested in interactions where feedback intensity is mapped to contact force between a virtual object and the fingerpads---it is a relocated actuator \cite{pezent2019tasbi}. 
It is important to note that wrist-worn devices are essentially relocated actuators when they are used for MR haptics: they stimulate the wrist without covering the hand. However, this means they inherit relocated actuators’ limitation: they cannot directly provide sensations in the hand, lowering the realism of feedback that users, when touching virtual UIs. To provide haptics that are felt in the hand from the wrist, one must utilize emerging approaches that propagate stimuli inside the body, which we will review next.
\vspace{-5 pt}
\subsection{Referred Sensations beyond Stimulation Points}
While creating haptic perception in the hand by stimulating the wrist sounds paradoxical, vibrations or electrical stimuli applied to the skin disperse inside the tissue. As such, researchers started exploring ways to propagate stimuli specifically to a target area, creating referred sensations---“somatosensory feelings that are perceived to emanate from a body part other than, but in association with, the body part being stimulated” \cite{mccabe2003referred}. One approach which has shown some recent progress, is to utilize constructive interference of vibrations: rendering haptics in the base of the finger with vibrations applied to the fingertip \cite{dandu2020rendering}; or creating a point of tactile sensation in the forearm with multiple actuators around that region \cite{de2023focused}. Unfortunately, we have yet to see this approach creating localized sensations within the hand/fingers from the wrist/arm.

\textbf{Referred sensations via electro-tactile feedback.} Electrical pulses applied to the skin can also create referred sensations by stimulating nerves under the skin that innervate tactile receptors in the hand \cite{micera2010wearable}. This has been primarily explored in neuroscience: stimulation applied to the forearm \cite{geng2012evaluation}, elbow \cite{forst2015surface}, and upper arm \cite{vargas2019evoked} to create sensations in the hand; or stimulating the lower palm to evoke sensations in the fingers \cite{d2017electro}. Building on these findings, researchers in haptics recently started developing haptic devices leveraging this phenomenon: electro-tactile feedback to the base of the finger \cite{ogihara2022transcutaneous, yoshimoto2011development} or the back of the hand \cite{tanaka2023full} to provide tactile sensations at the fingerpads. This supports both manual tasks with physical objects \& providing feedback for virtual UIs. Most relevant to our proposal, excitingly, electrical stimulation to the wrist can also create a coarse tactile sensation in the palm \cite{pena2021channel}, or even multiple areas of the palm by stimulating different regions of the wrist \cite{ogihara2023multi}. Although these results are appealing, this approach currently comes with critical limitations.
\vspace{-5 pt}
\subsection{Challenges in Electro-Tactile Feedback to the Wrist}
As reviewed, electro-tactile feedback may become a viable solution for generating tactile sensations in the hand via a wrist-worn device. However, two critical challenges remain that hinder its wider deployment in MR haptics.

\textbf{Perceived locations of the sensations.} As electrode placement moves away from the fingers toward the wrist, the nerves that innervate the fingers become more bundled \cite{tanaka2023full}. This poses a challenge when targeting a specific hand region with electrodes attached to the wrist. As Ogihara et al.’s recent findings demonstrate, sensations tend to spread across the palm and multiple fingers, making it difficult to isolate and target individual fingers, such as the index finger \cite{ogihara2023multi}. When considering adding feedback to UI elements like buttons and sliders in MR, it is essential for users to feel the sensation in their fingers, particularly in the most commonly used ones---the thumb and index finger. Unfortunately, the localizability of electro-tactile feedback is limited in this regard.

\textbf{Use of gel electrodes.} Existing electro-tactile devices for this approach use gel electrodes---a standard medical apparatus for improving skin conductivity. Unfortunately, this invalidates the devices’ practicality: gel electrodes are adhesive, hindering any detachment once they are attached \cite{tanaka2023interactive}, and they require frequent replacement as their conductivity drops within a few hours \cite{alba2010novel}. While we see the potential to enhance the haptic realism by electrically stimulating the wrist, ironically, it sacrifices the main advantages of wrist-worn haptics, that is, device practicality and social acceptability---much like how there is no sticky watch straps.

\section{Our Approach to Practical MR Wrist-Haptics}
We address the aforementioned challenges of wrist-applied electro-tactile feedback and turn it into a practical wrist-worn haptic interface for MR. Our approach is two-fold.
\vspace{-5 pt}
\subsection{Visually Augmented Electro-Tactile Feedback}
We discovered a novel perceptual illusion in which visually highlighting a target finger enhances the perceived localization of the electro-tactile feedback on that finger. Our approach draws from cross-modal illusions---a method to modify the haptic perception of virtual objects (e.g., shape \cite{ban2012modifying}, weight \cite{samad2019pseudo}) by modulating their appearance or movement. This includes altering the appearances of a user’s hand avatar in MR to modulate perceived haptic compliance \cite{morimoto2023effect, punpongsanon2015softar}. Unlike these works, our visually augmented electro-tactile feedback modulates the perceived locations of tactile sensations.

\textbf{Principle.} When visual and tactile stimuli are applied concurrently in proximity, the perceived location of the tactile stimulus gravitates toward the visual cue, a phenomenon known as “visuotactile ventriloquism”, which previously was shown with vibrations on the arm \cite{samad2016visual}. Extending this, our MR application visually highlights a target finger while our wrist-worn device delivers a tactile sensation to the hand. Thus, with visuotactile ventriloquism, the visual feedback shifts the perceived tactile sensation closer to the targeted finger. In our Study 2, we found that our approach creates about 50 and 40\% of perceived sensations within the thumb and index finger respectively---a significant improvement from electro-tactile alone. Importantly, as Samad et al. found, visuotactile ventriloquism diminishes as the visual and tactile stimuli get spatially disparate \cite{samad2016visual}. Consequently, for the fingers, this augmentation is uniquely feasible with electro-tactile feedback, which can mitigate the disparity by creating sensations in the hand. Conversely, vibrotactile or mechanical stimuli may not be able to produce the same effect. This is because these stimuli only generate sensations at the wrist, making the large disparity between visual and tactile feedback. This was confirmed in our pilot experiments and by the feedback from participants in Study 3.
\vspace{-5 pt}
\subsection{A Practical Wristband with Custom Dry Electrodes}
We engineered a custom electro-tactile wristband featuring dry electrodes. Switching from standard gel electrodes to dry electrodes allowed us to significantly improve the wearability and practicality of the device by eliminating the gel’s adhesiveness and the need for frequent replacement. While this switching might seem straightforward, naïvely removing the gels leads to issues as common metallic materials (e.g., copper) oxidize with the user’s skin and sweat, causing skin irritation and material degradation \cite{hostynek2004skin}. To mitigate this, we fabricated a synthetic conductive elastomer that is robust against oxidation (i.e., high bio-compatibility) and maintains the same level of conductivity as copper (see Section 4.2). Additionally, we customized our stimulus waveform to further mitigate the risk of irritating sensations from the electrodes (see Section 4.3). These material and stimulus designs enable our device to reliably deliver a pleasant “tapping” sensation in the hand without relying on gel electrodes, which we confirmed in our Study 1.

\subsection{Application Walkthrough}
\begin{figure}[H]
 \centering
 \includegraphics[width=0.95\columnwidth]{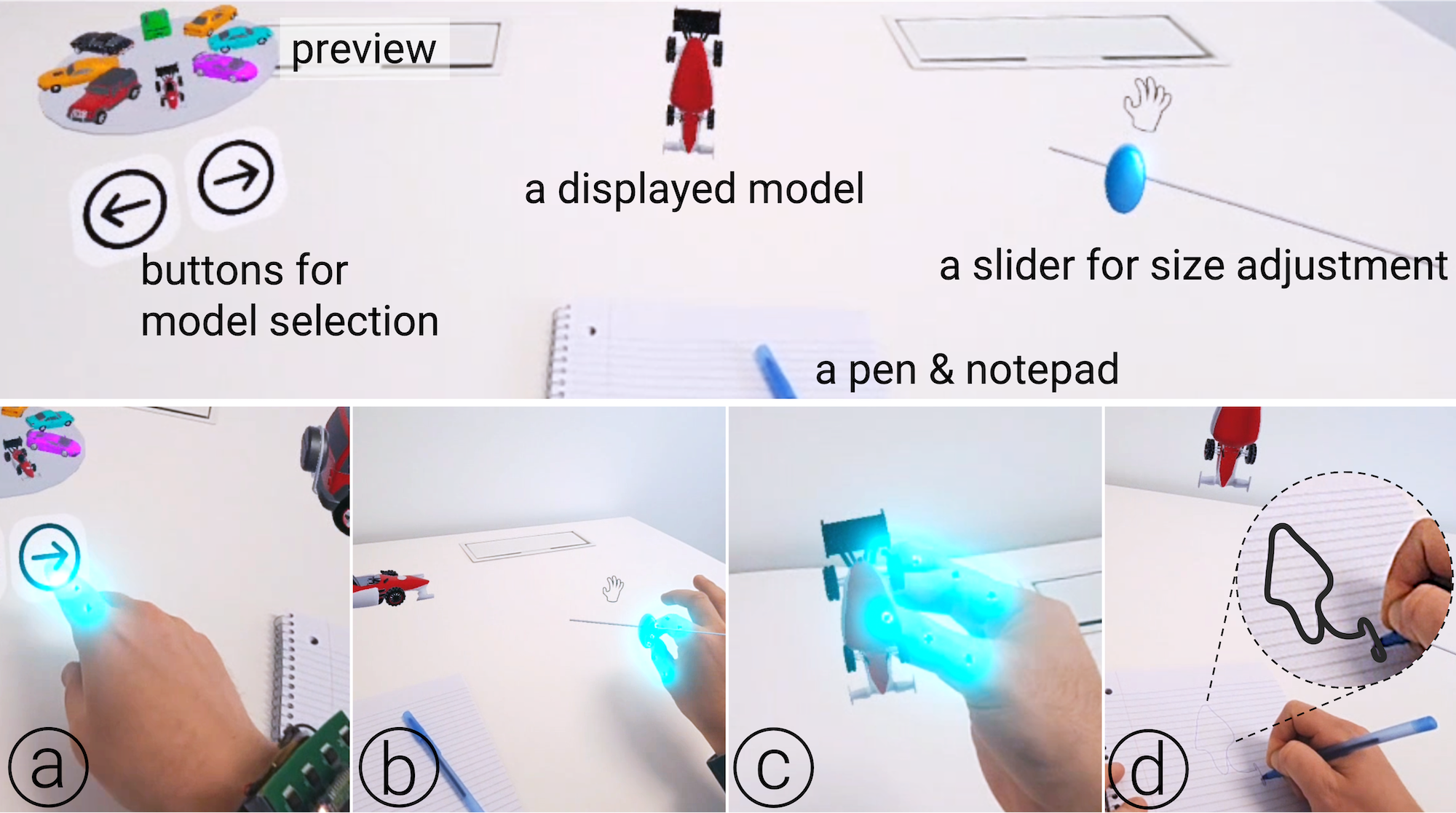}
 \vspace{-10 pt}
 \caption{A 3D model browser that embodies our tactile feedback.}
 \label{fig:app}
\end{figure}

Now, we walk through how our approach supports haptic interactions in practice. For proof of concept, we created a 3D model browser (\cref{fig:app}). The user feels the tactility of every single button press while browsing through different car models (\cref{fig:app}a). As they move the slider UI’s knob to adjust the scale of the model being selected, they feel tactile detents that represent discrete values (\cref{fig:app}b). The user can also directly grab the model to change its orientation, feeling the tactility of the object (\cref{fig:app}c). Finally, as our device leaves the hand completely free, the user can pick up a pen and sketch the model for their creative process (\cref{fig:app}d).
\vspace{-5 pt}
\vspace{-15 pt}
\section{Design and Implementation}
Here, we describe technical details of our electro-tactile wristband and stimuli, as well as visual \& tactile feedback control, that altogether compose our practical solution to MR haptics.

\subsection{Wristband Design and Electronics}
As shown in \cref{fig:teaser} (f), the main component of our device is the electrode strap that houses silicon-elastomer dry electrodes (17.8 $\times$ 3.8 $\times$ 1.2 cm; 47 g). The elastic band going through the electrode strap enables users to easily wear the device and adjust its tightness. The circuit board for electro-tactile stimulator, Bluetooth, battery, etc. (5.2 cm $\times$ 6 cm $\times$ 2 cm; 37 g) sits atop the user’s wrist (the dorsal side) when the device is worn.

\begin{figure}[H]
 \centering
 \includegraphics[width=0.95\columnwidth]{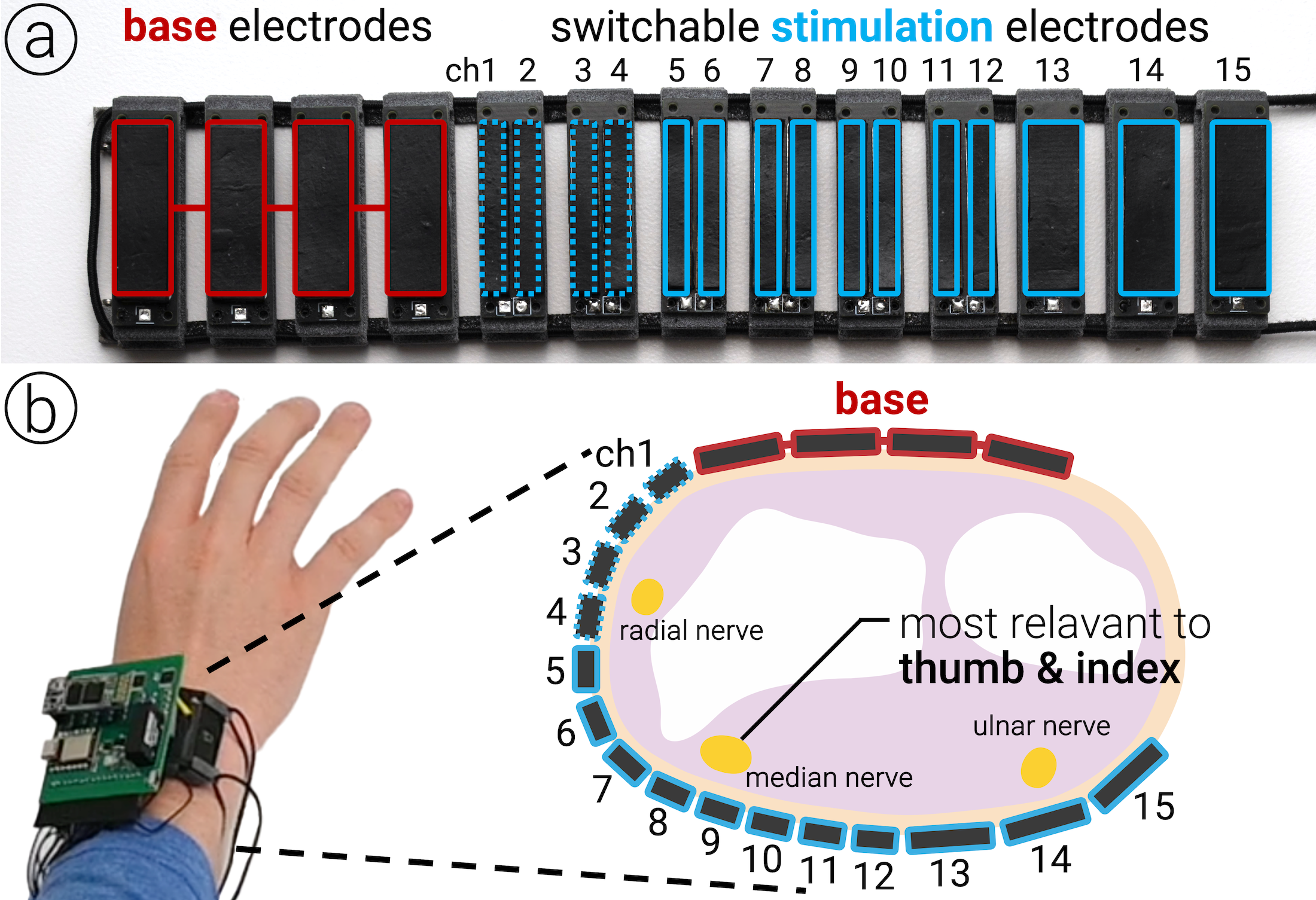}
 \vspace{-10 pt}
 \caption{(a) The configuration of the electrodes. (b) The arrangement of the electrodes around the wrist.}
 \label{fig:channel}
\end{figure}

\begin{figure}[H]
 \centering
 \includegraphics[width=0.8\columnwidth]{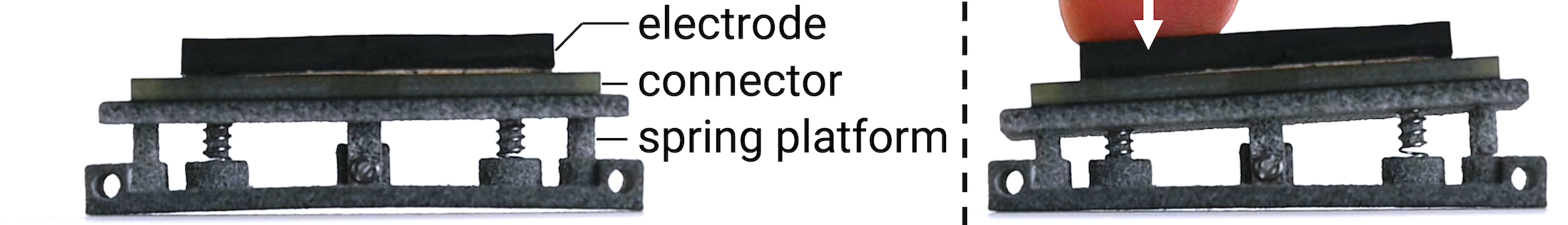}
 \vspace{-10 pt}
 \caption{Our spring unit allows the electrodes to tilt by $\pm$6 degrees}
 \label{fig:mechanical}
\end{figure}

\begin{figure}[H]
 \centering
 \includegraphics[width=0.95\columnwidth]{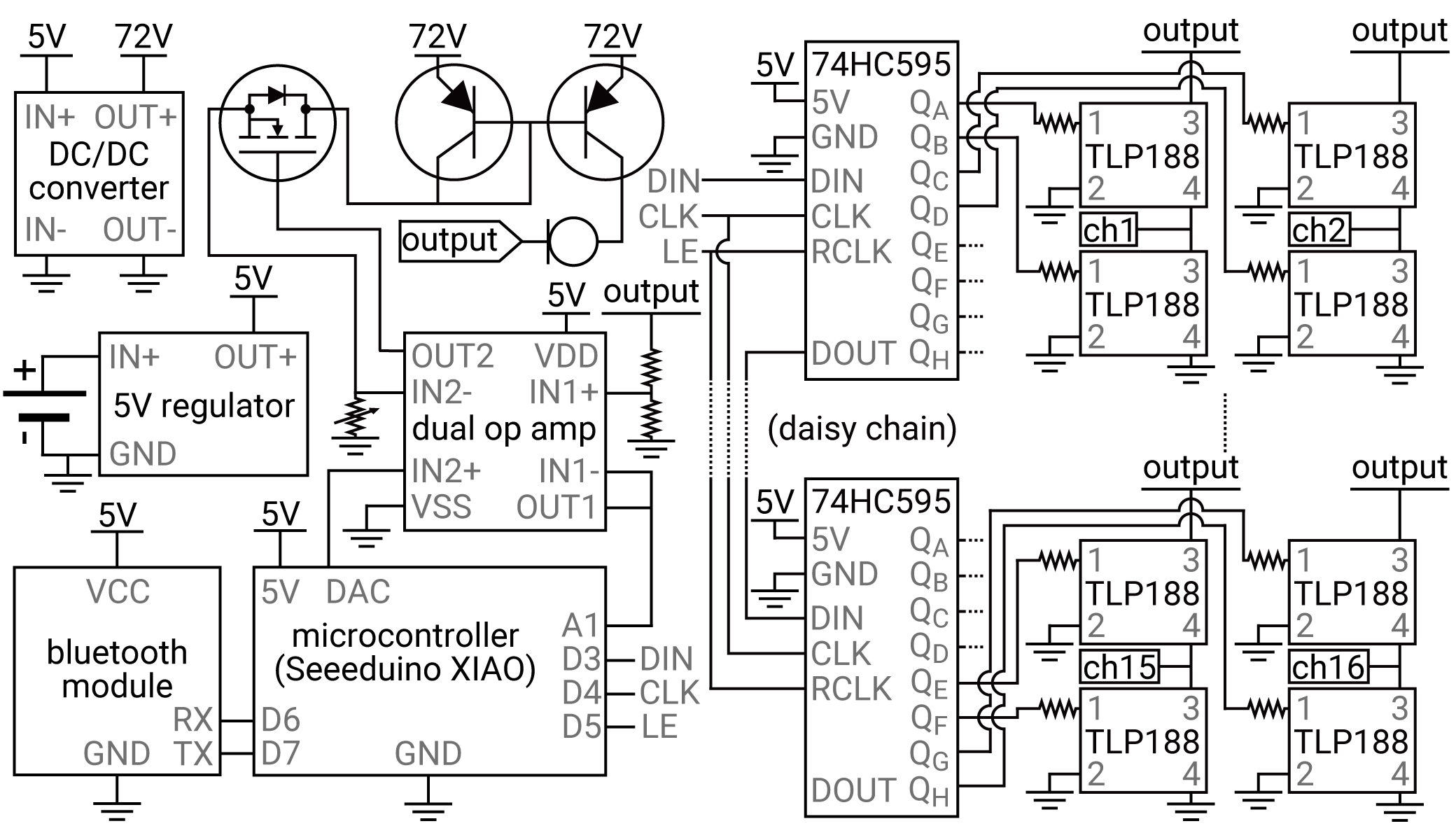}
 \vspace{-10 pt}
 \caption{A schematic diagram of our circuit board.}
 \label{fig:circuit}
\end{figure}

\textbf{Stimulation channels.} \cref{fig:channel} (a) shows the flipside of our electrode strap, consisting of 19 dry electrodes: base electrodes and switchable stimulation electrodes denoted as \textit{ch1} to \textit{ch15}. The electrical currents flow between one stimulation electrode (e.g., \textit{ch5}) and the four base electrodes (constant; fixed). Thus, by activating different stimulation electrodes, our device can stimulate different regions around the wrist. Note that base electrodes are not always the ground due to our stimulation waveform (see Section 4.3). As shown in \cref{fig:channel} (b), the stimulation electrodes lie over the nerves innervating receptors populated in the palmar side of the hand (i.e., median and ulnar). Since we primarily target the thumb \& index finger, we designed the electrodes over the median nerve to be smaller and densely distributed, enabling finer adjustment of the stimulation point. In our studies, we did not use the four electrodes highlighted with dashed blue lines (i.e., \textit{ch1$\sim$4}) as they would only stimulate the back of the hand (i.e., the radial nerve) as informed by prior work \cite{ogihara2023multi}.
\vspace{-5 pt}
\vspace{-5 pt}
Our electrode strap consists of 13 mechanical units; \cref{fig:mechanical} shows an individual unit. Each component embeds springs and allows the electrode to tilt by $\pm$6 degrees. This allows the band to conform to the user’s wrist---a surface with a complex curvature.
\vspace{-5 pt}
\vspace{-5 pt}
\textbf{Electronics.} \cref{fig:circuit} shows a detailed schematic of our stimulator. The stimulator is powered by a 3.7V LiPo battery with a 5V regulator (U1V11F5). A DC/DC converter (NMT0572) boosts up the 5V to a 72V line, which is used for the stimulation. Upon receiving a stimulation command via a HC-06 Bluetooth module, the microcontroller (Seeeduino XIAO) sends a pulse waveform (0$\sim$3.3V) to a voltage controlled current source consisting of an op-amp (LMV358) and a FET (BSS87). The converter maps the input to a load-independent current (0$\sim$4 mA). The output pulse further passes through a current mirror --- a pair of transistors (FCX705) --- and a current limiting diode (E-452). Finally, our 16-channel bipolar switching circuit assigns the output pulse to the target pair of electrodes. The switching circuit consists of eight daisy-chained shift registers (74HC595) that control 32 photorelays (TLP188).
\vspace{-5 pt}
\vspace{-5 pt}
\textbf{Safety.} Our circuit employs three levels of safety measures: (1) it controls electrical currents independent of skin resistance; (2) the current limiting diode caps the maximum current to 4.5 mA; and (3) it measures skin resistance in real-time via the voltage divider connected to the stimulation output.

\subsection{Elastomer-Based Dry \& Durable Electrodes}
\begin{figure}[H]
 \centering
 \includegraphics[width=0.95\columnwidth]{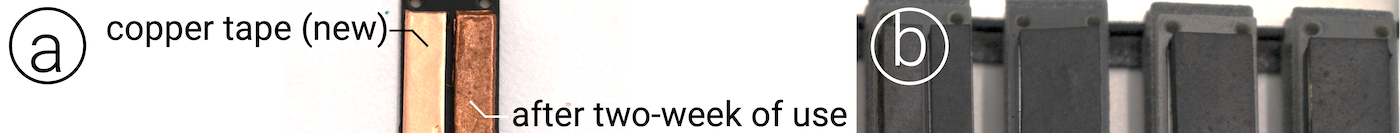}
 \vspace{-10 pt}
 \caption{(a) The degradation of copper-tape electrodes. (b) ours.}
 \label{fig:material}
\end{figure}

While replacing standard hydrogel with dry electrodes might sound simple, naively attaching off-the-shelf copper electrodes to the skin leads to a bio-compatibility issue. As shown in \cref{fig:material} (a), the copper-tape electrodes chemically react with the skin and oxidizes, resulting in skin irritation and, over time, material corrosion \cite{hostynek2004skin}. Therefore, we employ custom-made conductive elastomer (\cref{fig:material}b), which is chemically unreactive to the skin \cite{chen2020mechanically, ji2023skin}. As with other conductive elastomer, this chemical stability prevents the electrodes from material degradation and maintains the conductivity over time, eliminating the need for replacement---they are essentially pieces of silicon rubber.
\vspace{-5 pt}
\vspace{-5 pt}
\textbf{Material properties.} Our electrodes measured the conductivity of 0.62 $\Omega$·cm via four-point colinear probing and the hardness of 55 Shore A via 1600 Type A Durometer. A skin impedance measurement with PalmSens4 in its AC sweeping mode (1 mA, 142 Hz) \cite{rusanen2021laboratory} also showed no significant difference between ours (72.3 k$\Omega$) and copper tapes (76 k$\Omega$)

\textbf{Fabrication.} Our electrodes are composed of a tin-cure silicone (MoldMax 10T), 4-6nm carbon nanotubes (CNTs) and PR-19 carbon nanofibers (CNFs), following a composition formula that is publicly available \cite{reese_high_2023}. We first add 10wt\% CNTs to an uncured silicone compound and mix at 800 RPM for 2 minutes, followed by a 10-minute resting period. Then, we add 10wt\% CNFs and mix at 800 RPM for 2 minutes. Afterwards, we heat-press the compound at 150℃ for 45 minutes in a 100$\times$100$\times$2 mm mold. Once cured, we laser-cut the compound into the target shapes (26$\times$8$\times$2 or 26$\times$4$\times$2 mm). Finally, we attach the electrodes to our device’s connector parts using conductive adhesive (Dowsil EC6601).

\subsection{Stimulus Design for Charge-Balancing}
\begin{figure}[H]
 \centering
 \includegraphics[width=0.95\columnwidth]{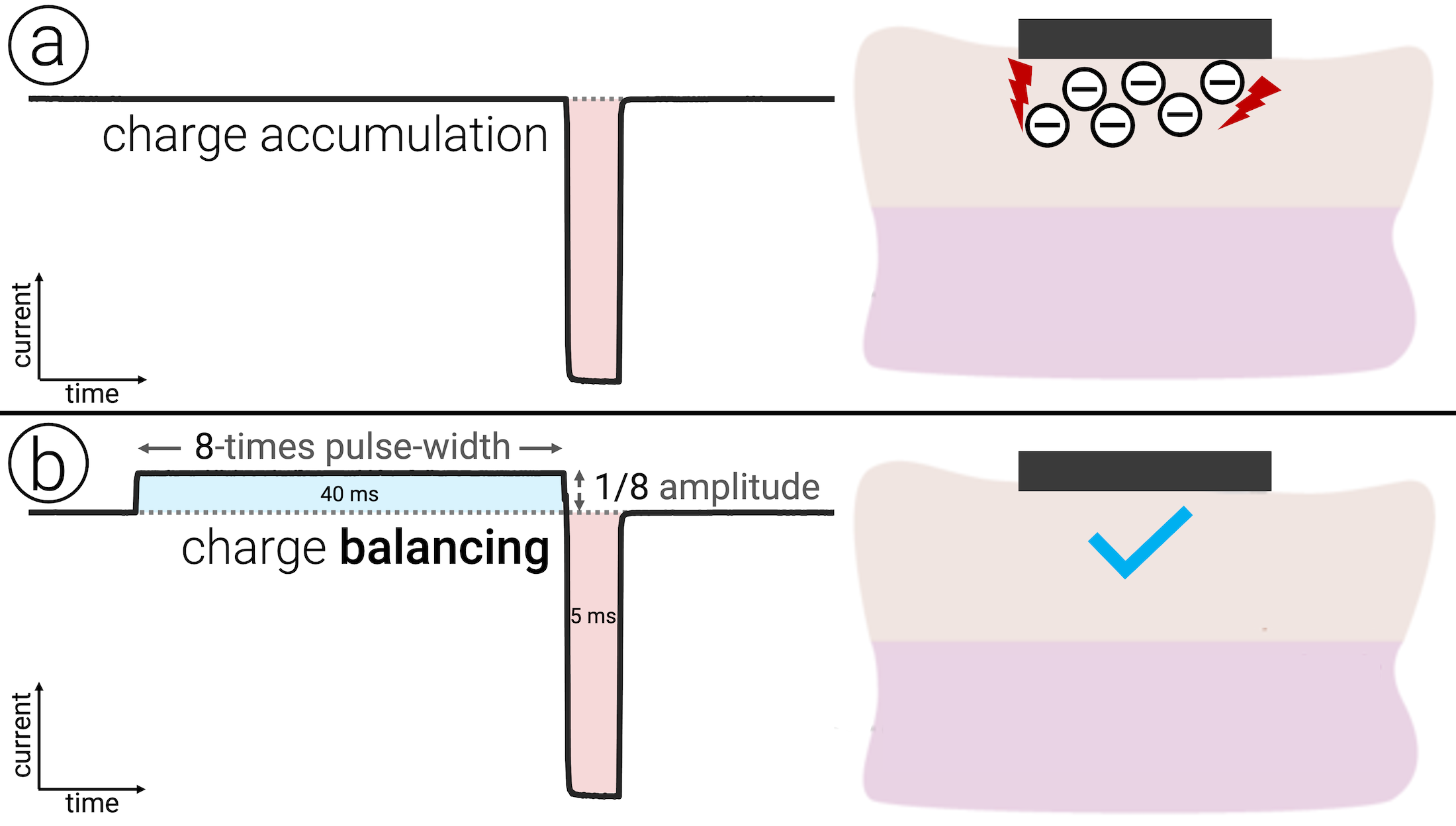}
 \vspace{-10 pt}
 \caption{(a) Conventional monophasic pulses (red). (b) We add an asymmetric priming pulse (blue), balancing the electron flow.}
 \label{fig:waveform}
\end{figure}

Beyond electrode materials, the stimulation itself is another factor that can irritate the skin. \cref{fig:waveform} (a) illustrates a pulse waveform commonly used for referred electro-tactile feedback, i.e., monophasic stimulation \cite{ogihara2023multi, tanaka2023full}.
\vspace{-5 pt}
\vspace{-5 pt}
Here, electrical current flows in one direction between the electrodes, building up a charge of electrons, which irritates the skin \cite{sooksood2010active}. One way to prevent this charging is to add a paired pulse with the same amplitude but opposite polarity, i.e., biphasic stimulation. This balances the electron flow between the two directions and thus prevents charge accumulation on the skin \cite{sooksood2010active}. Unfortunately, pulses of the opposite polarity (i.e., anodic stimulation) produce additional tactile sensations directly underneath the electrodes \cite{kajimoto2016electro}, in this case, at the wrist. This is undesirable as sensations in the wrist could distract users from the finger-oriented feedback.

This is why our stimulation waveform is asymmetric(\cref{fig:waveform}b): similar to biphasic stimulation, each 5-ms cathodic pulse (red) is paired with a 40-ms priming pulse (blue) of the opposite polarity. This means that, compared to the stimulation pulse, the priming pulse has only one-eighth the amplitude, but its pulse width is eight times longer. Thus, while balancing out the overall charge to prevent skin irritation, the lower amplitude of the priming pulse generates minimal tactile sensations at the wrist. As characterized in our Study 1, this stimulation design enables our device to predominantly create “tapping” sensations in the hand.

\subsection{Design of Visual Effects}
We carefully designed the color and size of our visual feedback. After testing multiple colors, we determined to adopt the light blue highlighting effects for their visibility and social acceptance. For the size, as users touch virtual UIs with their fingerpads, one may think that the visual effect should be applied only there. However, since electro-tactile feedback can only generate sensations over a larger area, just highlighting the fingertip leads to a spatial disparity between the visual and tactile stimuli. Unfortunately, as discussed by Samad et al., the large spatial disparity reduces the effect of the cross-modal illusion \cite{samad2016visual}. On the contrary, enlarging the visual feedback size to match the electro-tactile feedback (e.g., from the fingertip to the wrist) results in a mismatch with the target area (i.e., the fingerpad) where the user should feel the sensation. 
\vspace{-5 pt}
\begin{figure}[H]
 \centering
 \includegraphics[width=\columnwidth]{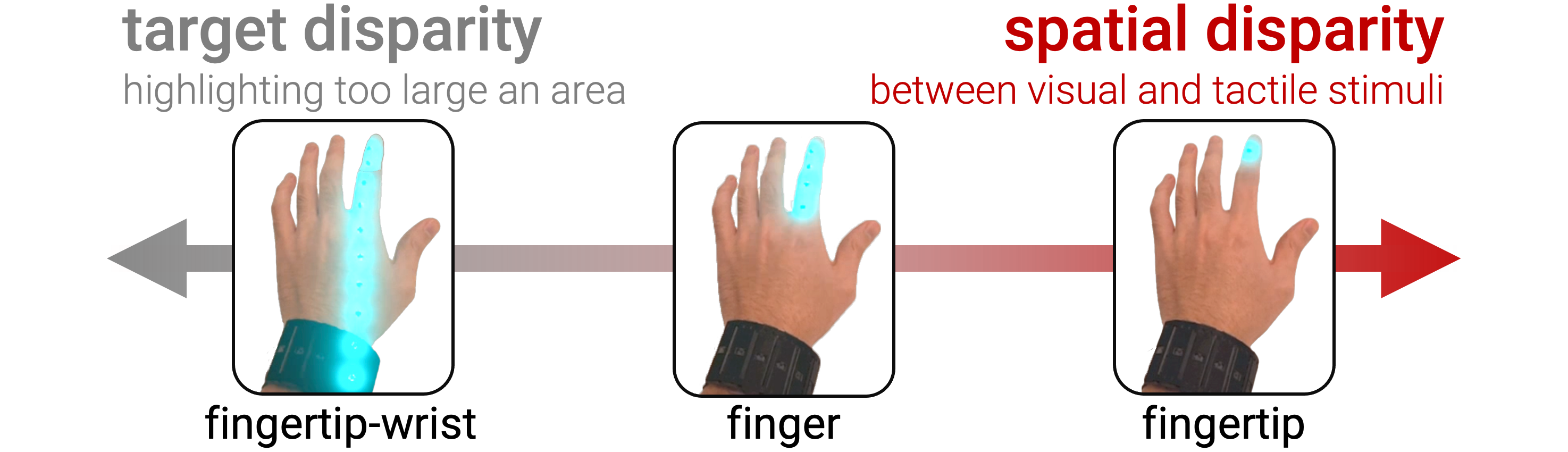}
 \vspace{-10 pt}
 \caption{The trade-off in the regions to apply visual effects.}
 \label{fig:design}
\end{figure}
\vspace{-5 pt}
After months of pilot exploration defining a design space, we sampled the following three sizes of visual effect and compared their perceived tactile sensations in our Study 2 (\cref{fig:design}): (1) highlighting the fingertip; (2) highlighting the finger; and (3) a highlight moving from the fingertip to the wrist. We found in Study 2 that the finger pattern, which balances the two extremes, results in better tactile localization in the target fingers. Consequently, we adopted it for our proposed applications and Study 3.

\subsection{Tracking, Display and Control}
\begin{figure}[H]
 \centering
 \includegraphics[width=0.95\columnwidth]{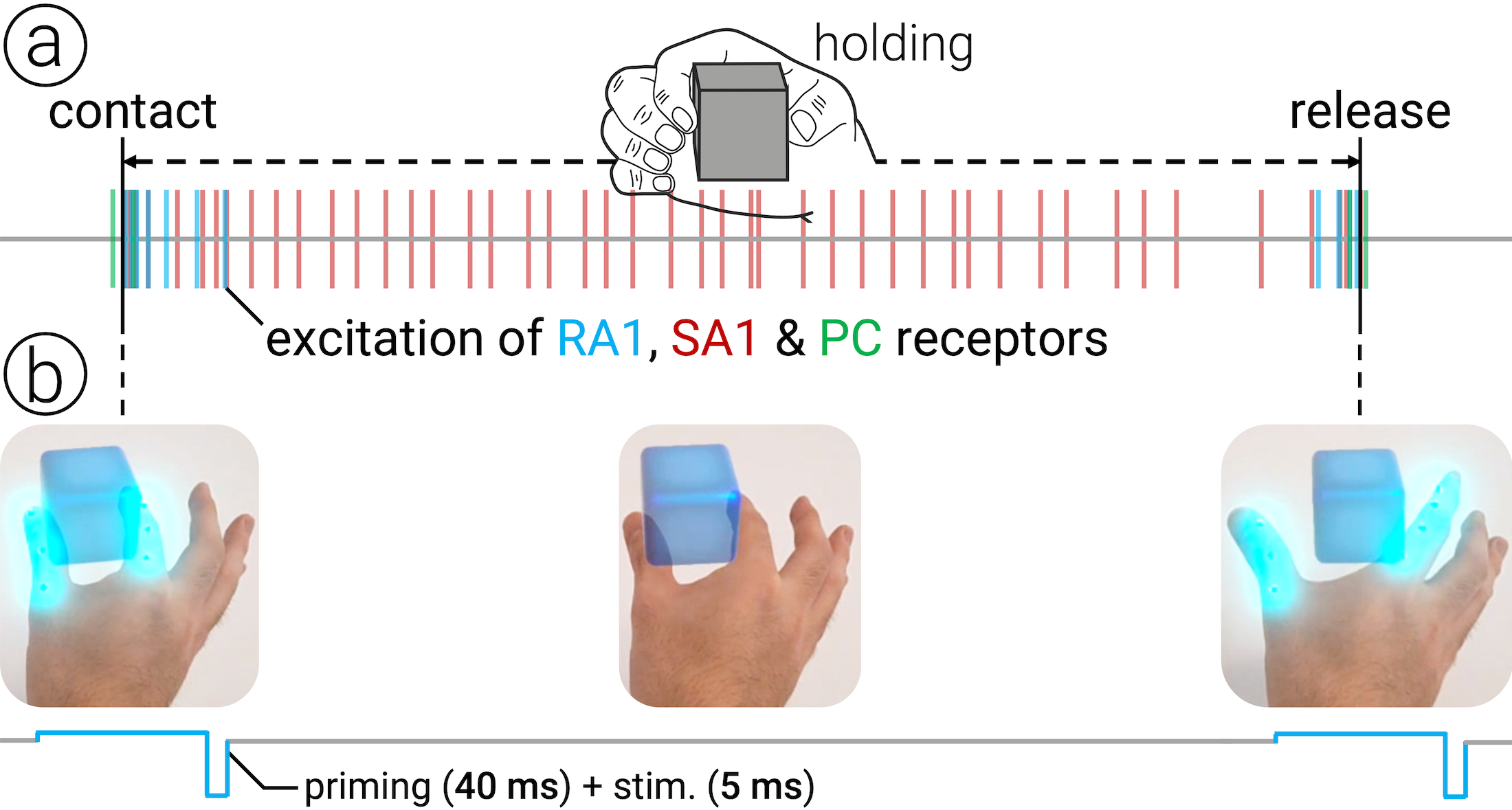}
 \vspace{-10 pt}
 \caption{(a) Excitation of RA1, SA1 \& PC tactile receptors in touching and grabbing a physical object (reproduced from \cite{kandel2000principles}). (b) Our control strategy of cathodic stimulation that reflects this principle.}
 \label{fig:control}
\end{figure}

Our MR applications run on a Quest 3 headset \& Unity3D, utilizing Quest’s PassThrough API. Based on the headset’s hand tracking, our system applies an electro-tactile stimulus concurrent with visual feedback upon contacting and releasing a virtual UI (\cref{fig:control}b). 

\vspace{-5 pt}
\vspace{-5 pt}

Note that the example shown is object-grabbing, but the principle also applies to other UIs such as buttons \& sliders. Our feedback mapping reflects the neurological basis of tactile perception: electrical stimuli activate axons of RA1 , SA1 and PC mechanoreceptors concurrently \cite{ogihara2023multi}---in touching or grabbing objects, the contact \& release are indeed the events that activate these three receptors at the same time \cite{johnson2001roles, kandel2000principles} (\cref{fig:control}a).

\section{Study 1: Perception with Electro-Tactile Alone}
This study focused on characterizing our electro-tactile feedback in isolation from visual feedback. While prior work evaluated locations of evoked sensations via stimulating the wrist \cite{ogihara2023multi}, their findings are specific to devices with gel electrodes and a sustained 2-second pulse duration. Our aim was to evaluate the efficacy of our dry electrode device and custom pulses in creating tactile cues in the fingers. This study was exempt from our Institutional Review Board (IRB) and passed our institution’s safety and privacy review.
\vspace{-5 pt}
\subsection{Study Design}
\textbf{Hypothesis, condition \& collected data.} Our hypothesis was that our device with the dry electrodes and custom short-pulses could still evoke referred sensations toward the thumb and index finger. We had a single condition: our electro-tactile feedback. We asked participants to illustrate where they felt the tactile sensation after stimulation---a conventional psychophysics method employed in this domain \cite{ogihara2023multi, scarpelli2020evoking, vargas2019evoked}. In addition, we asked them to describe the quality of sensation by selecting from the following five keywords: “tapping”, “vibrating”, “tingling”, “pressing”, or “skin-stretching”.

\textbf{Participants.} We recruited 12 participants from our institution (9 identified as male, 3 as female; 31.3 $\pm$ 5.2 years old; all were right-handed). Each study session took about 30 minutes.

\textbf{Apparatus.} Seated at a desk, participants wore our device on their non-dominant arm resting on an armrest, palm facing down (as in most MR touch interactions). This kept their dominant hand free for using our GUI (as in \cite{ogihara2023multi}). Since the tactile sensitivities of dominant \& non-dominant hands are similar \cite{van1997lack}, we assumed the study results would generalize to both hands. Our device’s circuit board was affixed to the desk, separate from the electrode strap, and connected via cables in accordance with our institution’s experimental safety protocol. On the participant’s dominant side, we provided an iPad showing our GUI \& an Apple Pencil for participants’ responses. The GUI showed an illustration of the palmar side of the hand (as in \cref{fig:study1}a) and the buttons for erasing the indication \& moving to the next trial. Additionally, for the current study, it had radio buttons with the labels (e.g., “tapping”) for reporting the quality of the sensation.

\textbf{Procedure.} Each participant performed 22 trials: 11 (ch5$\sim$ch15) stimulation channels $\times$ 2 repetitions. Note that the presentation order of the stimulation channel was randomized. In each trial, after a random waiting period, our device output ten stimulations, each followed by a one-second interval. Each stimulus consisted of a 40-ms priming pulse and a 5-ms stimulation pulse (see Section 4.3). Then, participants indicated the perceived area of the sensation and its strongest point on their palmar side of the hand using our GUI. We also asked them to report sensations outside of that side, if any, which we did not observe during the study. Finally, participants chose the quality of sensation from the five keywords.

\textbf{Calibration.} At the beginning of each trial, we performed calibration so that our system adapts the stimulation intensity for each finger (i.e., channel). We increased the current amount by 0.1 mA increments, ensuring it remained pain-free (the maximum current limit was set to 4 mA). We stopped at the intensity where participants noticed clear tactile sensation (i.e., threshold intensity). Moreover, prior to the first trial, we ensured that the device’s position on the participant’s wrist was aligned as depicted in \cref{fig:channel} (b), i.e., the front end of the device on the head of the ulnar bone and the first “base” electrode right next to the ulnar bone.

\textbf{Data analysis.} To characterize how well our device can localize the sensations when the stimulation channels were calibrated for the targets, we analyzed the participants’ responses for the two channels that have the highest rate of perceived tactile sensation in the thumb and index finger respectively (i.e., calibrated channels). 

\subsection{Results}
From the calibration, we found the average intensity of 0.65 mA (SD=0.26) for the thumb and 0.76 mA (SD=0.27) for the index.
\vspace{-5 pt}
\begin{figure}[H]
 \centering
 \includegraphics[width=0.95\columnwidth]{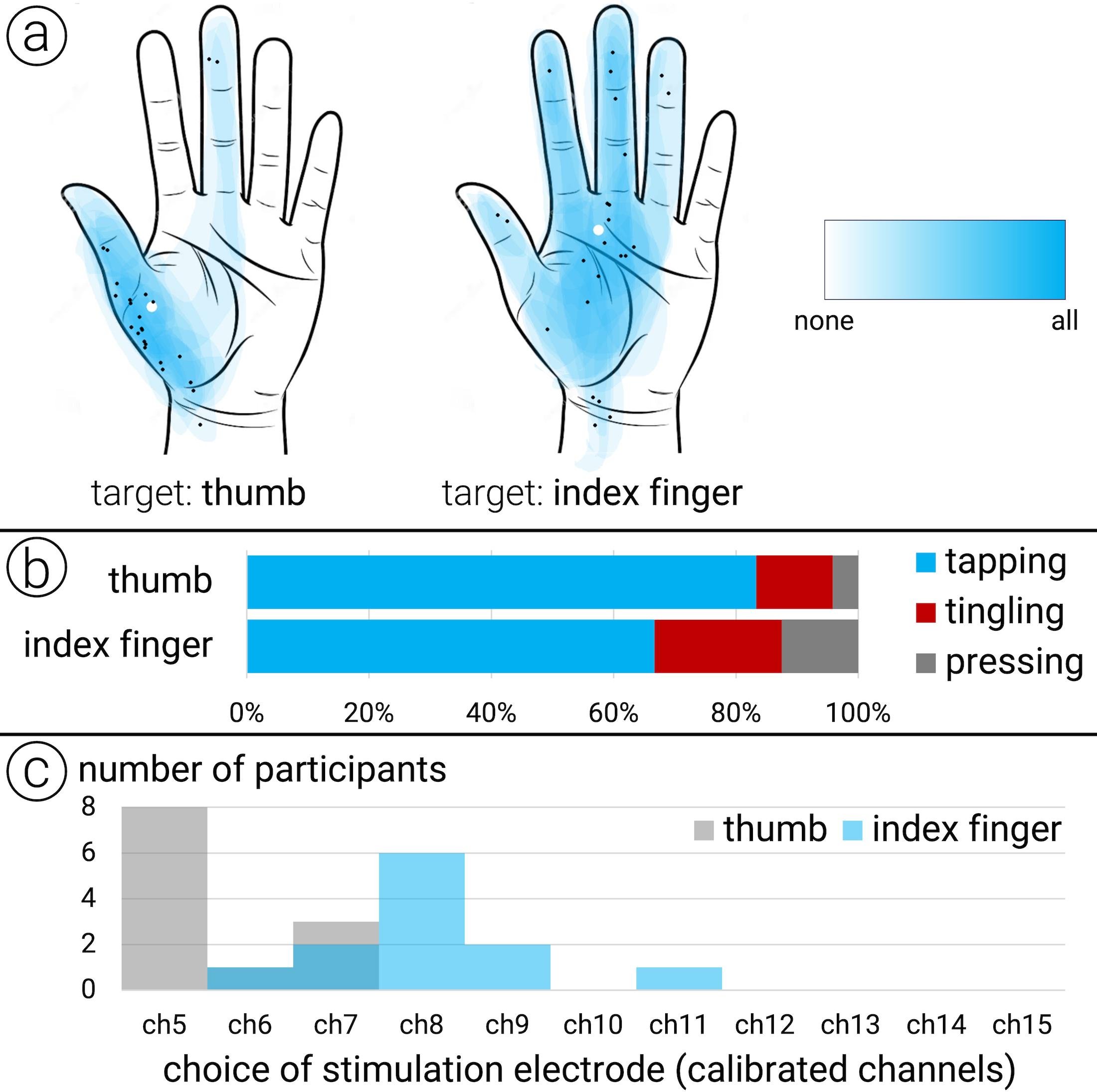}
 \vspace{-10 pt}
 \caption{(a) A heatmap of the perceived tactile sensations, compiled by overlaying raw data from all participants (black dots: the points of strongest sensation as reported by participants; white dots: their average coordinates). (b) Quality of the sensations. (c) Distribution of the calibrated channels.}
 \label{fig:study1}
\end{figure}
\vspace{-5 pt}
\cref{fig:study1} (a) displays a heatmap of the perceived sensations for all participants, when stimulating each individual with their calibrated channels. Aggregating over all participants, we found that 33.1\% (SD=28.8) of the elicited tactile sensation was felt in the thumb; in contrast, 16.1\% (SD=23.2) was felt in the index finger. For the strongest points of the sensation, 25\% of them (6 out of 24) were perceived within the thumb, and only 4\% (1 out of 24) were within the index finger.

\cref{fig:study1} (b) reveals that, for both target fingers, the quality of the induced sensation was predominantly described as “tapping,” followed by “tingling” and then “pressing.”

\cref{fig:study1} (c) illustrates which channels of our device were most effective for the target fingers. Looking at the distributions of the participants, there is a clear majority consensus: ch5 for the thumb, and ch8 for the index finger. Conversely, channels 12 to 15 were found to be not relevant to eliciting tactile sensation in these fingers. These results align with prior work \cite{ogihara2023multi} and the anatomical positioning of the nerves at the wrist.

Finally, these results are based on aggregating all participants’ responses---one way to communicate inter-subject diversity. To allow others to further explore this diversity, we have uploaded all the raw data from our study as supplementary material.

\section{Study 2: Perception with Visual Augmentation}
This study tackled our main research question: whether adding visual effects to our electrical stimuli could improve the perceived localization of sensation. To evaluate this, we employed the same study protocol as Study 1 but with the addition of different mixed-reality visual feedback to the electro-tactile feedback. This study was exempt from our Institutional Review Board (IRB) and passed our institution’s safety and privacy review.

\subsection{Study Design}
\textbf{Hypothesis, condition \& collected data.} Our hypothesis was that adding visual effects to our electrical stimuli would improve the perceived localization of sensation. To investigate the effect of visual effect design (see Section 4.4), we had three visual effect sizes depicted in \cref{fig:design}, with two levels of visual opacity (i.e., full \& half), totaling six conditions. As in Study 1, we asked participants to indicate the perceived locations of the sensations, omitting the reporting of their quality.

\textbf{Participants.} The same 12 participants took part in this study immediately after Study 1.

\textbf{Apparatus.} We employed the same setup as Study 1 with the addition of a Meta Quest 3 headset, which displayed our visual feedback. Note that the headset was set to the passthrough mode, allowing the participants to see the real-world environment.

\textbf{Procedure.} Each participant performed 24 trials: 6 visual feedback conditions $\times$ 2 target fingers $\times$ 2 repetitions. Note that the trial order was randomized. For this study, calibration was unnecessary as stimulation intensities and channels for the thumb \& index finger were informed by Study 1 and participants wore the device continuously between the studies. Each study session took about 30 minutes. In each trial, after a random waiting period, our device output ten 45-ms pulses, each followed by a one-second interval (see Section 5.1); each stimulus was coupled with visual feedback overlayed on the participant’s hand through the MR headset. Finally, the participant indicated the area of their perceived tactile sensation and its strongest point on our GUI.

\subsection{Results}

\begin{figure}[H]
 \centering
 \includegraphics[width=0.95\columnwidth]{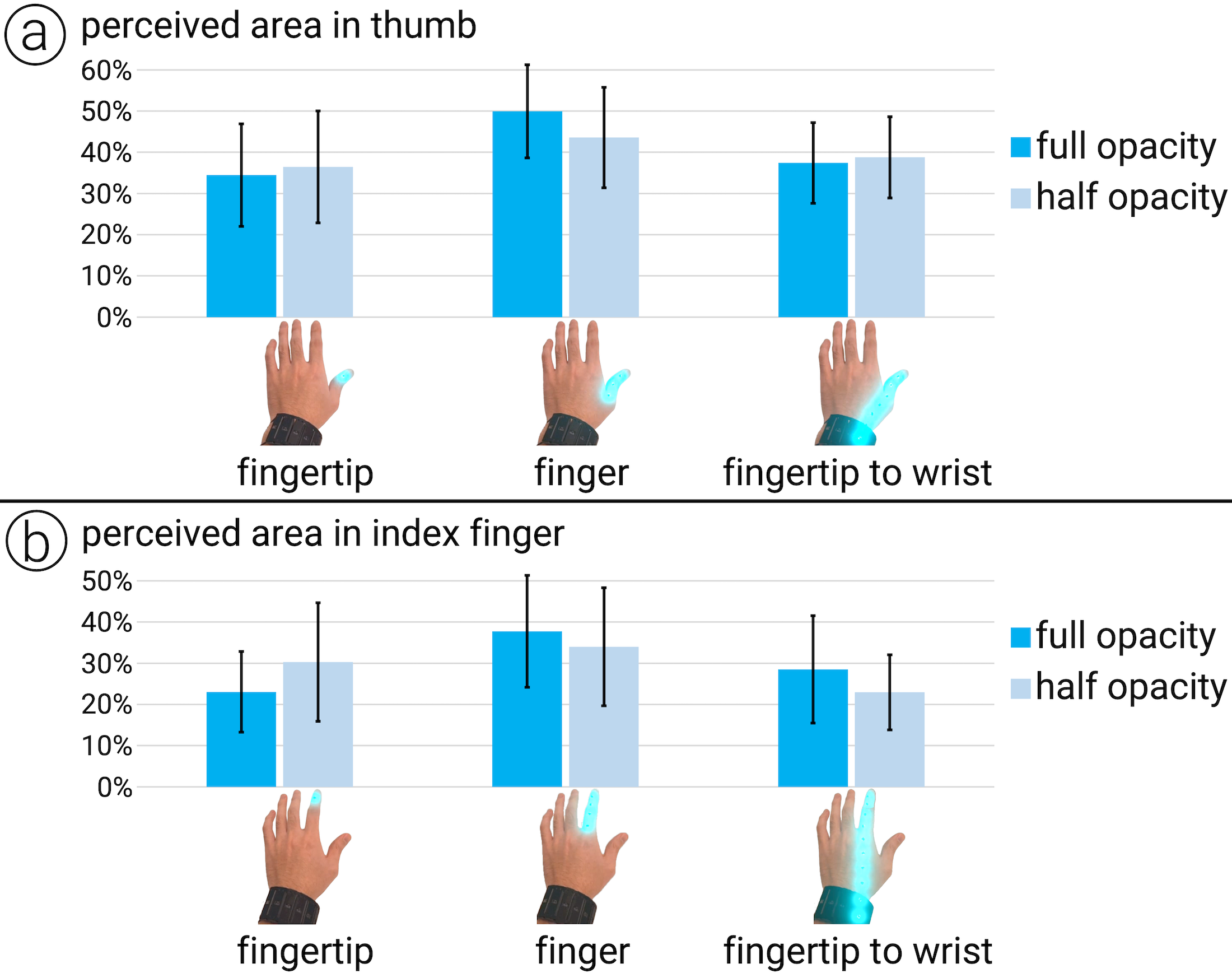}
 \vspace{-10 pt}
 \caption{The rate of tactile sensations in the target finger (a: thumb; b: index finger). The error bars show 95\% confidence interval.}
 \label{fig:study2main}
\end{figure}

\cref{fig:study2main} shows our main result; the rate of tactile sensation perceived in the target fingers during electrical stimulation with each of our six visual conditions.

\vspace{-5 pt}
\vspace{-5 pt}

\textbf{The effect of visual feedback design.} To understand the effect of the visual effect size and opacity, we conducted a two-way repeated measures ANOVA per target finger, having confirmed the sphericity via Mauchly’s test. For the thumb, we found the main effect of the visual effect size (F(2, 22)=5.3, p=0.01, ${\eta_p}^2$=0.03). We did not find main effects for the opacity (F(1, 11)=0.2, p=0.65, ${\eta_p}^2$=0) or for the interaction (F(2, 22)=1.3, p=0.28, ${\eta_p}^2$=0.01). We followed with a post-hoc test with Bonferroni corrections on the main effect of visual effect size and found significant differences between the full-finger and fingertip conditions (p$<$0.01), as well as between the full-finger and the fingertip-to-wrist conditions (p$<$0.01). Similarly, for the index finger, we found the main effect for the visual effect size (F(2, 22)=4.7, p=0.02, ${\eta_p}^2$=0.02). We did not find main effects for the opacity (F(1, 11)=0.1, p=0.81, ${\eta_p}^2$=0) or for the interaction (F(2, 22)=2.0, p=0.16, ${\eta_p}^2$=0.01). Similarly, we followed with a post-hoc test with Bonferroni corrections on the main effect of visual effect size and found significant differences between the full-finger and fingertip conditions (p$<$0.05), as well as between the full-finger and the fingertip-to-wrist conditions (p$<$0.01).

\vspace{-5 pt}
\begin{figure}[H]
 \centering
 \includegraphics[width=\columnwidth]{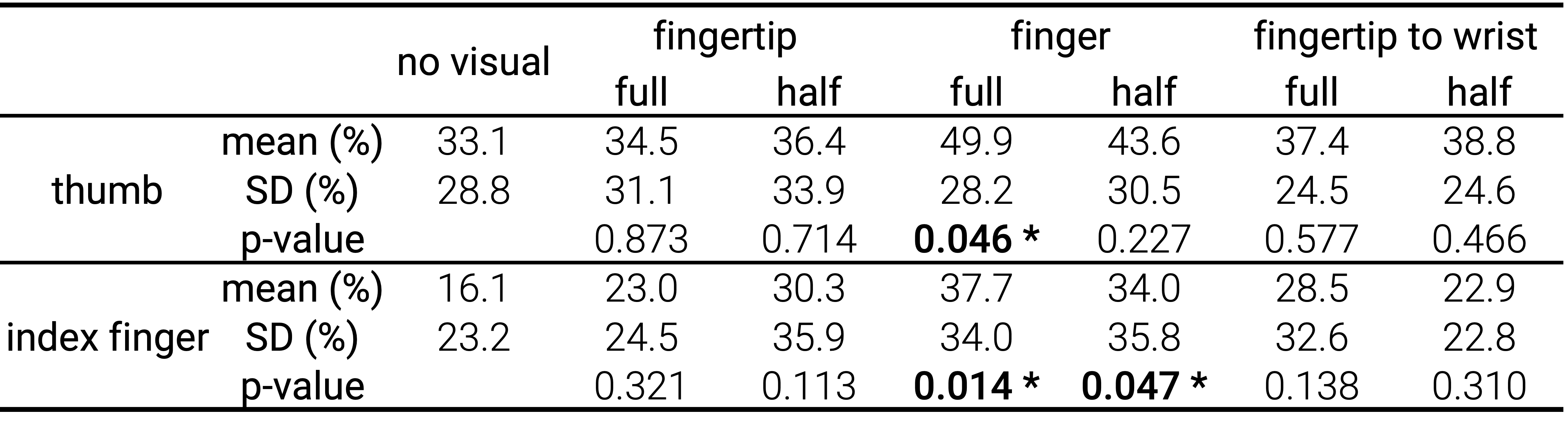}
 \vspace{-10 pt}
 \caption{(A summary table of unpaired t-tests comparing the rate of sensation felt in the finger from the six visual conditions to the no- visual condition (from Study 1).}
 \label{fig:table}
\end{figure}
\vspace{-5 pt}

\textbf{Vs. electro-tactile feedback alone (no visual).} We further analyzed the rate of tactile sensations felt in the target fingers via our augmented electro-tactile feedback in comparison to that via electro-tactile feedback alone (Study 1 results). \cref{fig:table} shows the results of an unpaired t test between the no visual condition (from Study 1) and each of the six visual feedback conditions. We observed that the visual augmentation applied to the finger with full opacity had a significantly higher rate of tactile sensations felt in the target finger, for both the thumb and index finger.

\begin{figure}[tb]
 \centering
 \includegraphics[width=0.95\columnwidth]{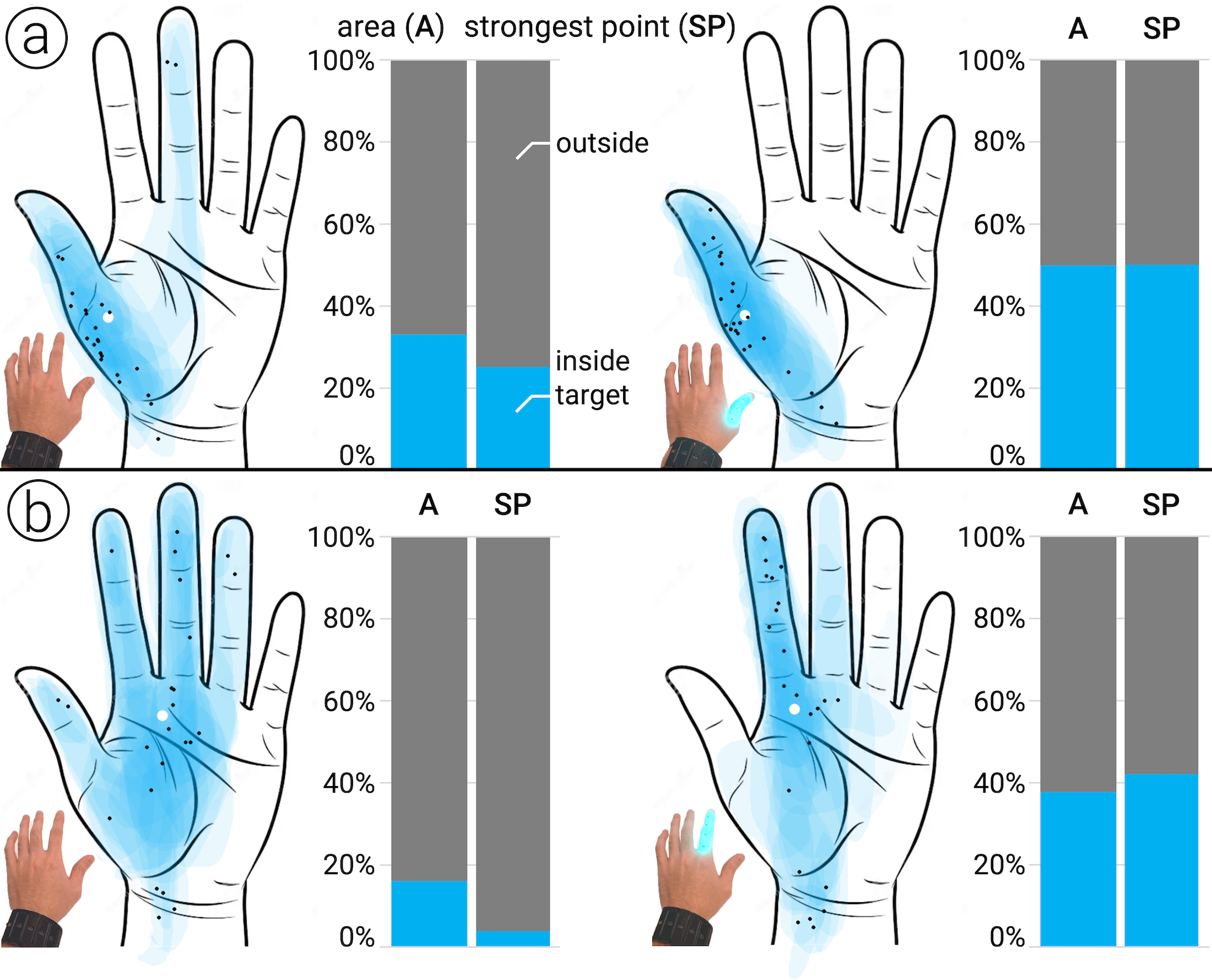}
 \vspace{-10 pt}
 \caption{The heatmaps of perceived sensations in the hand, aggregated for all participants for both the no-visual condition (from Study 1) and full-finger, full-opacity visual augmentation condition (from Study 2).}
 \label{fig:study2comparison}
\end{figure}

For further comparison between the with-visuals (finger-type) and without-visuals conditions, we plotted the heatmaps of the participants’ response side by side in \cref{fig:study2comparison}. Overall, this plot illustrates the comparison between the two conditions in terms of the localization of perceived tactile sensations in the target fingers. Note that, as per our study design, the electro-tactile feedback used in the two conditions are the same. For the thumb, the visual augmentation increased the ratio of perceived sensation from 33.1\% (SD=28.8) to 49.9\% (SD=28.2). For the index finger, the visual augmentation increased the ratio from 16.1\% (SD=23.2) to 37.7\% (SD=34). As for the strongest point of the sensation, with the visual augmentation, the participants indicated it in the target finger for 50\% of all trials (12 out of 24 cases) for the thumb; and for 42\% of all trials (10 out of 24 cases) for the index finger.

\textbf{Study conclusion.} Overall, we confirmed that visual augmentation could significantly improve the perceived localization of electro-tactile feedback. With the visual augmentation, we were able to create sensations that were perceived as occurring $\sim$50\% in the thumb and $\sim$38\% in the index.
\vspace{-5 pt}
\vspace{5 pt}
\section{Study 3: Haptic Realism in Mixed Reality}
Now we turn to understanding experiential aspects of our tactile feedback in comparison to a conventional vibrotactile wristband. As such, we had participants experience two MR interactions (button-pressing \& pinch-grabbing) with the two interface conditions. This study was exempt from our Institutional Review Board (IRB) and passed our institution’s safety and privacy review.

\subsection{Study Design}
\textbf{Hypothesis, condition \& collected data.} Our hypothesis was that our visually augmented electro-tactile feedback increases the haptic realism compared to traditional vibrotactile feedback. We had two conditions: ours and a baseline vibrotactile wristband (\cref{fig:study3}), each paired with our full-finger, full-opacity visual augmentation. We asked participants to rate their haptic experience in 7-point Likert scale for each condition and interviewed them regarding their experience (see Procedure).

\textbf{Participants.} We recruited eight participants from our institution (5 identified as male, 3 as female; 29.8 $\pm$ 4.2 years old; all were right-handed). Two had partaken in our previous study.

\textbf{Apparatus.} Participants wore a Quest 3 headset and a haptic device on their dominant wrist—either the electro-tactile or vibrotactile wristband. As with Study 1\&2, for both devices, their electronics were separate from the wristband components and affixed to the desk. Additionally, we provided a pen and paper for the questionnaires used in our evaluation.

\textbf{Vibrotactile wristband.} We used a linear resonant actuator (Vybronics, VLV152564W) positioned on the dorsal side of the wrist via a custom wristband mount. The actuator was driven by an audio amplifier (Nuvoton NAU8325) and emitted a sine wave at its resonant frequency (80 Hz) for 25 ms—similar to how Tasbi controls its vibrotactile feedback \cite{pezent2019tasbi}.

\textbf{Tasks.} The primary tasks were two mixed-reality interactions same as the ones depicted in \cref{fig:app} (a) \& (c) with different object appearance (\cref{fig:study3}a): (1) pressing a virtual button with the index finger; and (2) grabbing a virtual cube by pinching. In each task, participants repeated these actions eight times, as the button/cube reappeared in a new location after release. In both tasks, participants received visually augmented haptic feedback upon contact with and release of the object (see Section 4.5). These interactions were selected because they are basic and commonly encountered in MR. Following each condition, participants assessed the haptic realism of their experience by completing a questionnaire with a pen and paper. This component was included to incorporate manipulation of physical objects (e.g., grasping the pen and writing), and to evaluate the impact of the wristband devices on the participants’ dexterity.

\textbf{Procedure.} Each participant performed 4 trials: 2 tasks $\times$ 2 conditions, with the order of conditions counterbalanced. While we always adopted the maximum rated input (AC 1.4V) for the vibrotactile condition, for the electro-tactile condition, we calibrated the channels and stimulation intensities to enable tactile perception in the thumb and index finger. In the calibration process, unlike our first study, we verbally inquired with the participants about which part of the hand they felt the tactile stimuli on to adjust the stimulation channel. We set the intensity 0.1 mA higher than the minimum intensity required to evoke sensation (as in Study 1). During the trials, we encouraged the participants to “think aloud” and express any thoughts they had about the experience. After each trial, we asked the participants, “How well did the haptics match your visual impression of the experience?” — referring to the perceived haptic realism — and instructed them to rate this on a 7-point Likert scale (1: not at all; 7: completely) using the provided pen and questionnaire. Subsequently, we interviewed them about their reasoning behind their ratings. After completing all four trials, we conducted a semi-structured exit interview, initially asking “Which device do you prefer in terms of its haptic feedback?”. Their responses were recorded on the questionnaire, followed by an interview about their reasons. Finally, we probed into the effect of wearing the devices around the wrist on their manual dexterity by asking, “How much did the devices encumber grabbing \& handwriting with the pen to fill out the questionnaire?”.

\subsection{Results}
\begin{figure}[H]
 \centering
 \includegraphics[width=0.95\columnwidth]{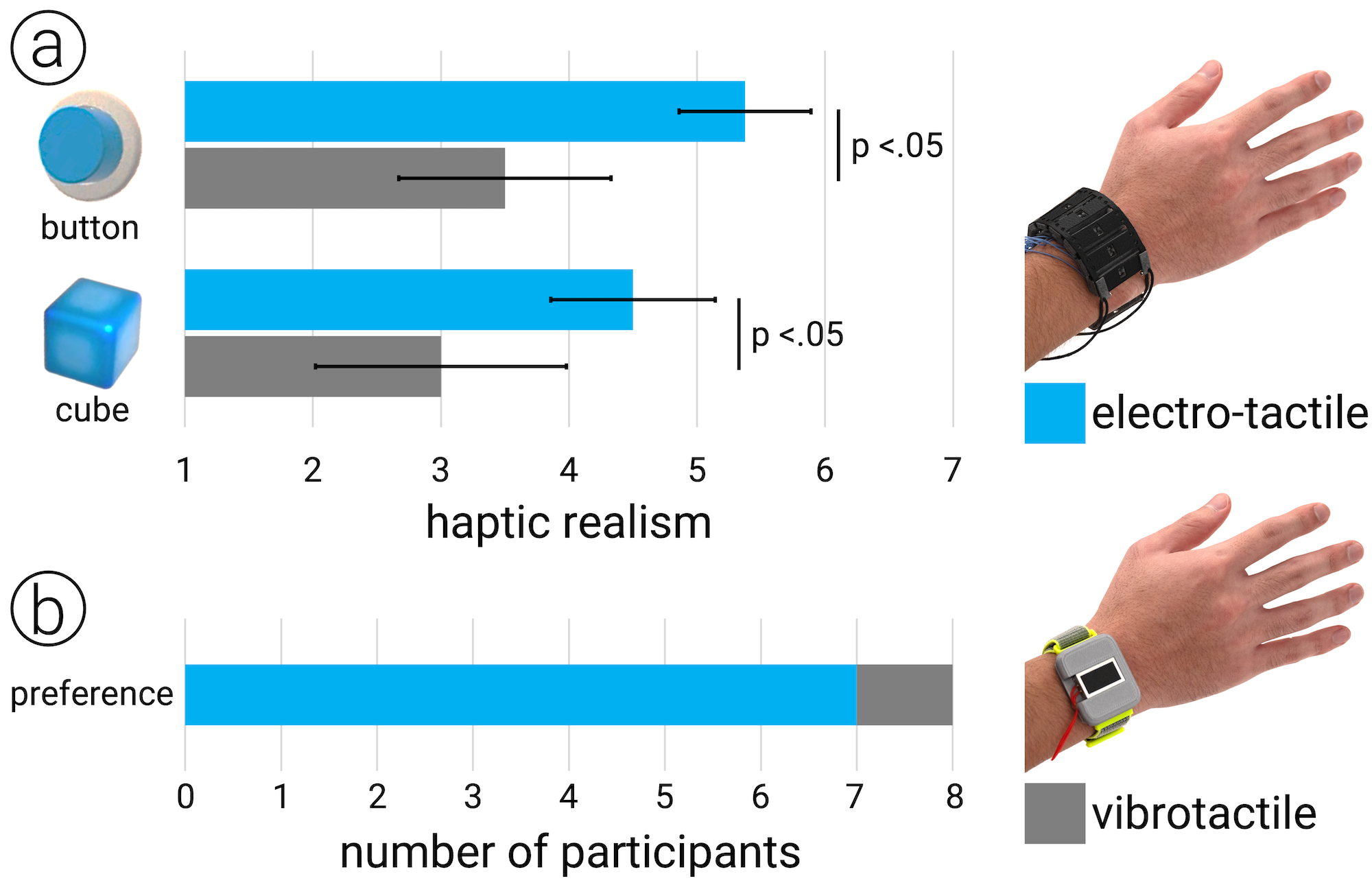}
 \vspace{-10 pt}
 \caption{(a) Ratings of haptic realism. Error bars indicate 95\% confidence intervals. (b) Participants’ preference on conditions.}
 \label{fig:study3}
\end{figure}

\cref{fig:study3} depicts the participants’ overall ratings of haptic realism for our electro-tactile feedback (button: M=5.4, SD=0.7; cube: M=4.5, SD=0.9) and the vibrotactile baseline (button: M=3.5, SD=1.2; cube: M=3.0, SD=1.4), respectively. Based on Wilcoxon signed-rank tests, we found a significant difference between the two conditions for both the button task (Z=-2.24, p=0.02) and the cube task (Z=-2.24, p=0.02). This suggests that participants experienced more realistic haptic feedback with our electro-tactile device. Moreover, seven out of eight participants preferred the electro-tactile device for haptic feedback after completing all trials. Now, we turn our attention to the participants’ comments.
\vspace{-5 pt}
\vspace{-5 pt}
\textbf{Feeling tactile sensations on the fingers vs. wrist.} Seven out of eight participants cited feeling tactile feedback toward their fingers as a key reason for rating our device’s haptic realism higher than the baseline. Comments included: “you can feel the perception on your palm and fingers (…) I think that’s a very big advantage” (P1); “I [had] the impact force [on] my finger” (P3); and “the haptic feedback is correctly [originating] from the finger” (P8). P1 further noted how the electro-tactile feedback aligned with the visual: “you see the visual like running from your fingertips down to the wrist and then also, somehow the electrical stimulation felt really like that (…) it was matching really well with the visual perception”. Conversely, for the vibrotactile device condition, five participants felt that the sensation occurring solely at the wrist was inconsistent with the interactions. Their comments included: “[the vibrotactile feedback] was definitely [in] the wrong place on the arm” (P6); and “(…) coming from the wrist so it feels wrong” (P8).

\textbf{Quality of the tactile sensations.} Four participants described how the sensation of electro-tactile feedback felt. Their comments included: “[it] gave the impression of having some impact force or interaction with a solid object” (P3); “[it] felt like the pressure” (P4); and “the haptic effect felt like a click and release (…) very much like I think about button presses feeling” (P6). In contrast, three participants commented on the sensation of the vibrotactile feedback. For instance, P1 remarked, “the vibration was so subtle”. P6 also stated, “[it] felt a lot less clicky and vaguer (…) like it was a buzz (…) didn’t feel as much like a button”.

\textbf{Encumberment from the devices.} Four participants mentioned the size of our electro-tactile wristband compared to the vibrotactile baseline: “the strap like one [was] heavier and stiffer”. (P3); and “the strap like device was thicker all the way around” (P6). However, upon asked about any encumbrance from the devices during the physical tasks, all participants noted that neither device directly obstructed their hands. Their comments included: “[wearing the wristbands felt] very natural” (P1); and “it’s not on the finger (…) I don’t think there will be any obstruction” (P5).

\textbf{Study conclusion.} Overall, qualitative feedback from participants suggests that our device provides more realistic feedback than the vibrotactile baseline. A key takeaway is that the visual augmentation did not enable the wrist-applied vibrotactile feedback to create tactile perception in the fingers. This can be interpreted in light of Samad et al.’s findings \cite{samad2016visual}: the visual augmentation diminishes as the spatial disparity in feedback (tactile vs. visual) increases.

\section{Limitation and Future Work}
Our approach is not without limitations: (1) our perceptual illusion is limited in MR, where visual effects are inherently available; (2) similar to most electrical stimulation techniques, ours requires calibration of stimulation channels and intensities for each user; (3) while it can localize sensations in the fingers, they are not strictly limited within the fingerpad regions; (4) our studies did not evaluate whether or how evoked tactile sensations might change over time; and (5) our studies had a small sample size ($\sim$12 participants), which was limited to right-handed young adults.

For future work, we are thrilled to expand the range of tactile sensations via our approach, exploring other stimulation parameters (e.g., pulse width, frequency). We are also excited to examine how visual feedback can be tailored to specific application contexts (e.g., electrical sparks for gaming, or even lower opacity levels) and how it affects our illusion. Finally, it would be valuable to evaluate the generalizability of our approach with a more diverse population.

\section{Conclusion}
We presented a novel method for delivering tactile feedback to the thumb and index finger from a wristband. By stimulating the user’s wrist while augmenting the stimulation with visual feedback over the target fingers, we were able to create about 40$\sim$50\% of the tactile sensations within the target fingers. This is the state-of-the-art in any wrist-worn devices that render touch toward the fingers. Moreover, we integrated our stimulation into a practical wristband design, replacing gelled electrodes typically required for electro-tactile stimulation with custom-made dry electrodes. In our series of user studies, we validated the effectiveness of our visually augmented electro-tactile feedback in creating sensations in the fingers. We also explored the haptic realism of our approach in comparison to a conventional vibrotactile wristband as a baseline, where participants preferred our approach to the baseline.

\acknowledgments{
We would like to thank Pornthep Preechayasomboon, Eric Whitmire, Wolf Kienzle, and Hrvoje Benko from Reality Labs Research for their support on this work. We are grateful to anonymous reviewers for their constructive feedback.
}

\bibliographystyle{abbrv-doi-narrow}

\bibliography{template}

\begin{thebibliography}{10}
\renewcommand*{\sfdefault}{PTSansNarrow-TLF}

\bibitem{noauthor_cybergrasp_nodate}
{CyberGrasp}.

\bibitem{alba2010novel}
N.~A. Alba, R.~J. Sclabassi, M.~Sun, and X.~T. Cui.
\newblock Novel hydrogel-based preparation-free eeg electrode.
\newblock {\em IEEE transactions on neural systems and rehabilitation engineering}, 18(4):415--423, 2010.

\bibitem{ando2002smartfinger}
H.~Ando, T.~Miki, M.~Inami, and T.~Maeda.
\newblock Smartfinger: nail-mounted tactile display.
\newblock In {\em ACM SIGGRAPH 2002 conference abstracts and applications}, pp. 78--78, 2002.

\bibitem{ban2012modifying}
Y.~Ban, T.~Kajinami, T.~Narumi, T.~Tanikawa, and M.~Hirose.
\newblock Modifying an identified curved surface shape using pseudo-haptic effect.
\newblock In {\em 2012 IEEE Haptics Symposium (HAPTICS)}, pp. 211--216. IEEE, 2012.

\bibitem{carcedo2016hapticolor}
M.~G. Carcedo, S.~H. Chua, S.~Perrault, P.~Wozniak, R.~Joshi, M.~Obaid, M.~Fjeld, and S.~Zhao.
\newblock Hapticolor: Interpolating color information as haptic feedback to assist the colorblind.
\newblock In {\em Proceedings of the 2016 CHI Conference on Human Factors in Computing Systems}, pp. 3572--3583, 2016.

\bibitem{chen2008tactor}
H.-Y. Chen, J.~Santos, M.~Graves, K.~Kim, and H.~Z. Tan.
\newblock Tactor localization at the wrist.
\newblock In {\em Haptics: Perception, Devices and Scenarios: 6th International Conference, EuroHaptics 2008 Madrid, Spain, June 10-13, 2008 Proceedings 6}, pp. 209--218. Springer, 2008.

\bibitem{chen2020mechanically}
S.~Chen, L.~Sun, X.~Zhou, Y.~Guo, J.~Song, S.~Qian, Z.~Liu, Q.~Guan, E.~Meade~Jeffries, W.~Liu, et~al.
\newblock Mechanically and biologically skin-like elastomers for bio-integrated electronics.
\newblock {\em Nature communications}, 11(1):1107, 2020.

\bibitem{dandu2020rendering}
B.~Dandu, Y.~Shao, and Y.~Visell.
\newblock Rendering spatiotemporal haptic effects via the physics of waves in the skin.
\newblock {\em IEEE Transactions on Haptics}, 14(2):347--358, 2020.

\bibitem{de2023focused}
V.~de~Vlam, M.~Wiertlewski, and Y.~Vardar.
\newblock Focused vibrotactile stimuli from a wearable sparse array of actuators.
\newblock {\em IEEE Transactions on Haptics}, 16(4):511--517, 2023.

\bibitem{d2017electro}
M.~D’Alonzo, L.~F. Engels, M.~Controzzi, and C.~Cipriani.
\newblock Electro-cutaneous stimulation on the palm elicits referred sensations on intact but not on amputated digits.
\newblock {\em Journal of neural engineering}, 15(1):016003, 2017.

\bibitem{forst2015surface}
J.~C. Forst, D.~C. Blok, J.~P. Slopsema, J.~M. Boss, L.~A. Heyboer, C.~M. Tobias, and K.~H. Polasek.
\newblock Surface electrical stimulation to evoke referred sensation.
\newblock {\em Journal of Rehabilitation Research \& Development}, 52(4), 2015.

\bibitem{geng2012evaluation}
B.~Geng, K.~Yoshida, L.~Petrini, and W.~Jensen.
\newblock Evaluation of sensation evoked by electrocutaneous stimulation on forearm in nondisabled subjects.
\newblock {\em Journal of Rehabilitation Research \& Development}, 49(2), 2012.

\bibitem{gupta2017hapticclench}
A.~Gupta, A.~A.~R. Irudayaraj, and R.~Balakrishnan.
\newblock Hapticclench: Investigating squeeze sensations using memory alloys.
\newblock In {\em Proceedings of the 30th Annual ACM Symposium on User Interface Software and Technology}, pp. 109--117, 2017.

\bibitem{han2018hydroring}
T.~Han, F.~Anderson, P.~Irani, and T.~Grossman.
\newblock Hydroring: Supporting mixed reality haptics using liquid flow.
\newblock In {\em Proceedings of the 31st Annual ACM Symposium on User Interface Software and Technology}, pp. 913--925, 2018.

\bibitem{hostynek2004skin}
J.~J. Hostynek and H.~I. Maibach.
\newblock Skin irritation potential of copper compounds.
\newblock {\em Toxicology Mechanisms and Methods}, 14(4):205--213, 2004.

\bibitem{ji2023skin}
H.~Ji, M.~Wang, Y.~Wang, Z.~Wang, Y.~Ma, L.~Liu, H.~Zhou, Z.~Xu, X.~Wang, Y.~Chen, et~al.
\newblock Skin-integrated, biocompatible, and stretchable silicon microneedle electrode for long-term emg monitoring in motion scenario.
\newblock {\em npj Flexible Electronics}, 7(1):46, 2023.

\bibitem{johnson2001roles}
K.~O. Johnson.
\newblock The roles and functions of cutaneous mechanoreceptors.
\newblock {\em Current opinion in neurobiology}, 11(4):455--461, 2001.

\bibitem{kajimoto2016electro}
H.~Kajimoto.
\newblock Electro-tactile display: principle and hardware.
\newblock {\em Pervasive Haptics: Science, Design, and Application}, pp. 79--96, 2016.

\bibitem{kandel2000principles}
E.~R. Kandel, J.~H. Schwartz, T.~M. Jessell, S.~Siegelbaum, A.~J. Hudspeth, S.~Mack, et~al.
\newblock {\em Principles of neural science}, vol.~4.
\newblock McGraw-hill New York, 2000.

\bibitem{kovacs2020haptic}
R.~Kovacs, E.~Ofek, M.~Gonzalez~Franco, A.~F. Siu, S.~Marwecki, C.~Holz, and M.~Sinclair.
\newblock Haptic pivot: On-demand handhelds in vr.
\newblock In {\em Proceedings of the 33rd Annual ACM Symposium on User Interface Software and Technology}, pp. 1046--1059, 2020.

\bibitem{liao2016edgevib}
Y.-C. Liao, Y.-L. Chen, J.-Y. Lo, R.-H. Liang, L.~Chan, and B.-Y. Chen.
\newblock Edgevib: Effective alphanumeric character output using a wrist-worn tactile display.
\newblock In {\em Proceedings of the 29th Annual Symposium on User Interface Software and Technology}, pp. 595--601, 2016.

\bibitem{mccabe2003referred}
C.~McCabe, R.~Haigh, P.~Halligan, and D.~Blake.
\newblock Referred sensations in patients with complex regional pain syndrome type 1.
\newblock {\em Rheumatology}, 42(9):1067--1073, 2003.

\bibitem{micera2010wearable}
S.~Micera, T.~Keller, M.~Lawrence, M.~Morari, and D.~B. Popovic.
\newblock Wearable neural prostheses.
\newblock {\em IEEE Engineering in Medicine and Biology Magazine}, 29(3):64--69, 2010.

\bibitem{morimoto2023effect}
K.~Morimoto, K.~Hashiura, and K.~Watanabe.
\newblock Effect of virtual hand's fingertip deformation on the stiffness perceived using pseudo-haptics.
\newblock In {\em Proceedings of the 29th ACM Symposium on Virtual Reality Software and Technology}, pp. 1--10, 2023.

\bibitem{moriyama2022wearable}
T.~Moriyama and H.~Kajimoto.
\newblock Wearable haptic device presenting sensations of fingertips to the forearm.
\newblock {\em IEEE Transactions on Haptics}, 15(1):91--96, 2022.

\bibitem{nittala2019like}
A.~S. Nittala, K.~Kruttwig, J.~Lee, R.~Bennewitz, E.~Arzt, and J.~Steimle.
\newblock Like a second skin: Understanding how epidermal devices affect human tactile perception.
\newblock In {\em Proceedings of the 2019 CHI Conference on Human Factors in Computing Systems}, pp. 1--16, 2019.

\bibitem{ogihara2022transcutaneous}
S.~Ogihara, T.~Amemiya, and K.~Aoyama.
\newblock Transcutaneous electrical nerve stimulation along the base of the finger to modify the location of tactile sensation at the finger.
\newblock In {\em 2022 IEEE International Symposium on Mixed and Augmented Reality Adjunct (ISMAR-Adjunct)}, pp. 646--647. IEEE, 2022.

\bibitem{ogihara2023multi}
S.~Ogihara, T.~Amemiya, H.~Kuzuoka, T.~Narumi, and K.~Aoyama.
\newblock Multi surface electrodes nerve bundles stimulation on the wrist: Modified location of tactile sensation on the palm.
\newblock {\em IEEE Access}, 11:13794--13809, 2023.

\bibitem{pasquero2011haptic}
J.~Pasquero, S.~J. Stobbe, and N.~Stonehouse.
\newblock A haptic wristwatch for eyes-free interactions.
\newblock In {\em Proceedings of the SIGCHI Conference on Human Factors in Computing Systems}, pp. 3257--3266, 2011.

\bibitem{pena2021channel}
A.~Pena, J.~Abbas, and R.~Jung.
\newblock Channel-hopping during surface electrical neurostimulation elicits selective, comfortable, distally referred sensations.
\newblock {\em Journal of neural engineering}, 18(5):055004, 2021.

\bibitem{pezent2019tasbi}
E.~Pezent, A.~Israr, M.~Samad, S.~Robinson, P.~Agarwal, H.~Benko, and N.~Colonnese.
\newblock Tasbi: Multisensory squeeze and vibrotactile wrist haptics for augmented and virtual reality.
\newblock In {\em 2019 IEEE World Haptics Conference (WHC)}, pp. 1--6. IEEE, 2019.

\bibitem{pohl2017squeezeback}
H.~Pohl, P.~Brandes, H.~Ngo~Quang, and M.~Rohs.
\newblock Squeezeback: Pneumatic compression for notifications.
\newblock In {\em Proceedings of the 2017 CHI Conference on Human Factors in Computing Systems}, pp. 5318--5330, 2017.

\bibitem{preechayasomboon2021haplets}
P.~Preechayasomboon and E.~Rombokas.
\newblock Haplets: Finger-worn wireless and low-encumbrance vibrotactile haptic feedback for virtual and augmented reality.
\newblock {\em Frontiers in Virtual Reality}, 2:738613, 2021.

\bibitem{punpongsanon2015softar}
P.~Punpongsanon, D.~Iwai, and K.~Sato.
\newblock Softar: Visually manipulating haptic softness perception in spatial augmented reality.
\newblock {\em IEEE transactions on visualization and computer graphics}, 21(11):1279--1288, 2015.

\bibitem{reese_high_2023}
S.~Reese, W.~PAN, R.~HUISZOON, Z.~Liu, P.~Wei, N.~Guo, M.~T. Khbeis, L.~Yao, A.~A. Albarran, K.~ERICKSON, F.~He, T.~J.~F. Wallin, and L.~Chen.
\newblock High aspect-ratio filler-containing conductive elastomeric materials and methods of use, Oct. 2023.

\bibitem{rusanen2021laboratory}
M.~Rusanen, S.~Myllymaa, L.~Kalevo, K.~Myllymaa, J.~T{\"o}yr{\"a}s, T.~Lepp{\"a}nen, and S.~Kainulainen.
\newblock An in-laboratory comparison of focusband eeg device and textile electrodes against a medical-grade system and wet gel electrodes.
\newblock {\em IEEE Access}, 9:132580--132591, 2021.

\bibitem{salazar2018path}
J.~Salazar, K.~Okabe, and Y.~Hirata.
\newblock Path-following guidance using phantom sensation based vibrotactile cues around the wrist.
\newblock {\em IEEE Robotics and Automation Letters}, 3(3):2485--2492, 2018.

\bibitem{samad2019pseudo}
M.~Samad, E.~Gatti, A.~Hermes, H.~Benko, and C.~Parise.
\newblock Pseudo-haptic weight: Changing the perceived weight of virtual objects by manipulating control-display ratio.
\newblock In {\em Proceedings of the 2019 CHI Conference on Human Factors in Computing Systems}, pp. 1--13, 2019.

\bibitem{samad2016visual}
M.~Samad and L.~Shams.
\newblock Visual--somatotopic interactions in spatial perception.
\newblock {\em Neuroreport}, 27(3):180--185, 2016.

\bibitem{scarpelli2020evoking}
A.~Scarpelli, A.~Demofonti, F.~Terracina, A.~L. Ciancio, and L.~Zollo.
\newblock Evoking apparent moving sensation in the hand via transcutaneous electrical nerve stimulation.
\newblock {\em Frontiers in Neuroscience}, 14:534, 2020.

\bibitem{sooksood2010active}
K.~Sooksood, T.~Stieglitz, and M.~Ortmanns.
\newblock An active approach for charge balancing in functional electrical stimulation.
\newblock {\em IEEE Transactions on Biomedical Circuits and Systems}, 4(3):162--170, 2010.

\bibitem{tanaka2023full}
Y.~Tanaka, A.~Shen, A.~Kong, and P.~Lopes.
\newblock Full-hand electro-tactile feedback without obstructing palmar side of hand.
\newblock In {\em Proceedings of the 2023 CHI Conference on Human Factors in Computing Systems}, pp. 1--15, 2023.

\bibitem{tanaka2023interactive}
Y.~Tanaka, A.~Takahashi, and P.~Lopes.
\newblock Interactive benefits from switching electrical to magnetic muscle stimulation.
\newblock In {\em Proceedings of the 36th Annual ACM Symposium on User Interface Software and Technology}, pp. 1--12, 2023.

\bibitem{teng2021touch}
S.-Y. Teng, P.~Li, R.~Nith, J.~Fonseca, and P.~Lopes.
\newblock Touch\&fold: A foldable haptic actuator for rendering touch in mixed reality.
\newblock In {\em Proceedings of the 2021 CHI Conference on Human Factors in Computing Systems}, pp. 1--14, 2021.

\bibitem{van1997lack}
A.~Van~Turnhout, J.~J. Hage, P.~De~Groot, and E.~Delange-De~Klerk.
\newblock Lack of difference in sensibility between the dominant and non-dominant hands as tested with semmes-weinstein monofilaments.
\newblock {\em The Journal of Hand Surgery: British \& European Volume}, 22(6):768--771, 1997.

\bibitem{varga_haptx_nodate}
S.~Varga.
\newblock {HaptX}: {Haptic} gloves for virtual reality and robotics.

\bibitem{vargas2019evoked}
L.~Vargas, G.~Whitehouse, H.~Huang, Y.~Zhu, and X.~Hu.
\newblock Evoked haptic sensation in the hand with concurrent non-invasive nerve stimulation.
\newblock {\em IEEE Transactions on Biomedical Engineering}, 66(10):2761--2767, 2019.

\bibitem{withana2018tacttoo}
A.~Withana, D.~Groeger, and J.~Steimle.
\newblock Tacttoo: A thin and feel-through tattoo for on-skin tactile output.
\newblock In {\em Proceedings of the 31st Annual ACM Symposium on User Interface Software and Technology}, pp. 365--378, 2018.

\bibitem{yoshimoto2011development}
S.~Yoshimoto, Y.~Kuroda, M.~Imura, and O.~Oshiro.
\newblock Development of a spatially transparent electrotactile display and its performance in grip force control.
\newblock In {\em 2011 Annual International Conference of the IEEE Engineering in Medicine and Biology Society}, pp. 3463--3466. IEEE, 2011.

\end{thebibliography}
\end{document}